\documentclass[11pt]{article}

\usepackage{epsfig}
\usepackage{psfig}
\def\laq{\raise 0.4ex\hbox{$<$}\kern -0.8em\lower 0.62
ex\hbox{$\sim$}}
\def\gaq{\raise 0.4ex\hbox{$>$}\kern -0.7em\lower 0.62
ex\hbox{$\sim$}}

\begin{document}

\begin{titlepage}
\begin{flushright}
CERN-PH-TH/2005-152
\end{flushright}

\vspace{0.8cm}

\begin{center}

\huge{Magnetized CMB anisotropies}
\vspace{0.8cm}

\large{Massimo Giovannini \footnote{e-mail address: massimo.giovannini@cern.ch}}

\normalsize
\vspace{0.3cm}
{{\sl Centro ``Enrico Fermi", Compendio del Viminale, Via 
Panisperna 89/A, 00184 Rome, Italy}}\\
\vspace{0.3cm}
{{\sl Department of Physics, Theory Division, CERN, 1211 Geneva 23, Switzerland}}
\vspace*{2cm}

\begin{abstract}
\noindent
Possible effects of large-scale magnetic fields on the Cosmic 
Microwave Background (CMB) are reviewed. Depending on the specific branch of the spectrum of plasma excitations, magnetic fields are  treated either within a two-fluid plasma description or within an effective (one-fluid) approach. The uniform field approximation is contrasted with the fully  inhomogeneous field approximation.  It is argued that the interplay between CMB physics and large-scale magnetic fields will represent a rather interesting cross-disciplinary arena along the next few years.
\end{abstract}
\end{center}

\end{titlepage}

\newpage

\renewcommand{\theequation}{1.\arabic{equation}}
\section{Why CMB anisotropies could be magnetized?}
\setcounter{equation}{0}

Simplified magneto-hydrodynamical estimates 
imply that  the magnetic diffusivity length scale\footnote{The 
magnetic diffusivity length is the typical scale 
below which magnetic fields are diffused because of the 
finite value of the conductivity of the (interstellar) medium.}
 in the interstellar medium is  of the order of the astronomical unit. 
On the other hand, magnetic fields are present over much larger 
length-scales so that there seems to be rather compelling 
evidence that galaxies, clusters and possibly super-clusters are 
all magnetized. The question that arises naturally in this 
context concerns the possible effects of large-scale 
magnetic fields on the Cosmic Microwave Background (CMB). 
In the last fifty years various cosmological mechanisms for the origin of large 
scale magnetic fields have been proposed. While different 
mechanisms rely on diverse physical assumptions some general features can be identified:
\begin{itemize}
\item{} the majority of the cosmological mechanisms imply 
the existence of large-scale magnetic fields after equality 
(but before decoupling);
\item{} the magnetic field present after equality 
is, according to the mentioned mechanisms, fully inhomogeneous;
\item{}  the typical amplitudes and length-scales 
of the magnetic field are characteristic of the given model.
\end{itemize}

The first possibility we can think of implies that magnetic fields 
are produced, at a given epoch in the life of the Universe,
inside the Hubble radius, for instance by a phase transition 
or by any other phenomenon able to generate a charge separation and, 
ultimately, an electric current.  In this context, the 
correlation scale of the field is much smaller that the typical scale 
of the gravitational collapse of the proto-galaxy which is 
of the order of the ${\rm Mpc}$. 
In fact, if the Universe is decelerating and if 
the correlation scale evolves as 
the scale factor, the Hubble radius grows  much faster 
than the correlation scale. 
Of course, one might invoke the possibility that the correlation 
scale of the magnetic field evolves more rapidly than the scale 
factor. A well founded physical rationale for this occurrence 
is what is normally called inverse cascade, i.e. the possibility 
that  magnetic (as well as kinetic) energy density is transferred 
from small to large scales. This implies, in real space, that 
 (highly energetic) small scale magnetic domains may coalesce
 to form magnetic domains of smaller energy but over larger 
 scales.  
 In the best of all possible situations, i.e. when inverse cascade 
 is very effective, it seems rather hard 
 to justify a growth of the correlation scale that would eventually 
 end up into a ${\rm Mpc}$ scale at the onset of gravitational collapse.
\begin{figure}
\centerline{\epsfxsize = 11cm  \epsffile{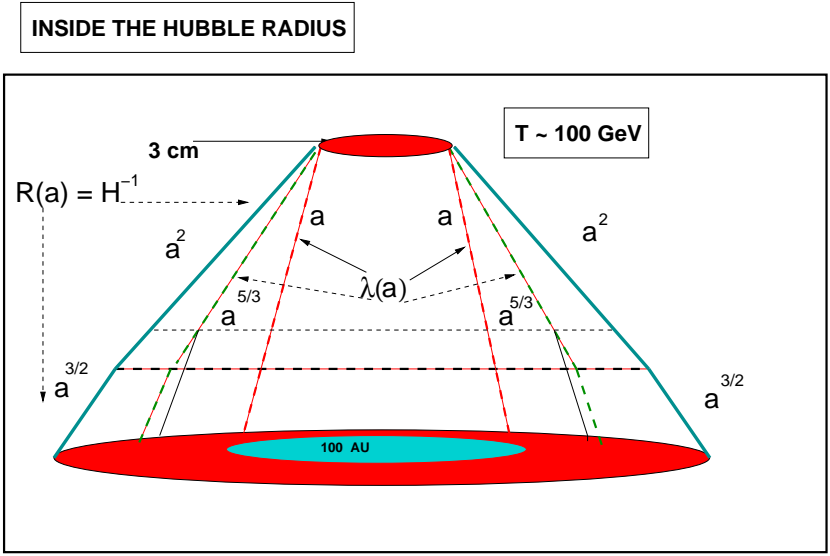}}
\vskip 3mm
\caption[a]{Evolution of the correlation scale 
for magnetic fields produced inside the Hubble radius. The horizontal thick dashed line marks the end of the radiation-dominated phase and the onset of the matter-dominated phase. The horizontal thin dashed line marks 
the moment of $e^{+}$--$e^{-}$ annihilation (see also footnoote 2). 
The full (vertical) lines represent the evolution of the Hubble radius 
during the different stages of the life of the Universe. The dashed (vertical) lines 
illustrate the evolution of the correlation scale of the magnetic fields. 
In the absence of inverse cascade the evolution of the correlation scale is given 
by the (inner) vertical dashed lines. If inverse cascade takes place the evolution 
of the correlation scale is faster than the first power of the scale factor (for instance
$a^{5/3}$) but always slower than the Hubble radius.}
\label{FIG1} 
\end{figure}
In Fig. \ref{FIG1} we report a schematic illustration of the evolution 
of the Hubble radius $R_{H}$ and of the correlation scale of 
the magnetic field as a function of the scale factor.  In Fig. \ref{FIG1} the 
horizontal dashed line simply marks the end of the radiation-dominated 
phase and the onset of the matter dominated phase: while above the dashed line 
the Hubble radius evolves as $a^2$ (where $a$ is the scale factor), below the dashed 
line the Hubble radius evolves as $a^{3/2}$. 

We consider, for 
simplicity, a magnetic field whose typical correlation scale is as 
large as the Hubble radius at the electro-weak epoch when the 
temperature of the plasma was of the order of $100\,\, {\rm GeV}$.
If the correlation scale evolves as the scale factor, the Hubble radius 
at the electroweak epoch (roughly $3$ cm) projects today over a scale 
of the order of the astronomical unit. If inverse cascades are invoked, the 
correlation scale may grow, depending on the specific features 
of the cascade, up to $100$ A.U.  or even up to $100$ pc. In both
 cases the final scale is too small if compared with the typical 
scale of the gravitational collapse of the proto-galaxy. In Fig. \ref{FIG1} 
a particular model for the evolution of the correlation scale $\lambda(a)$ has 
been reported \footnote{Notice, as it will be discussed later, that the inverse 
cascade lasts, in principle, only down to the time of $e^{+}--e^{-}$ 
annihilation (see also thin dashed horizontal line in Fig. \ref{FIG1}) since for temperatures smaller than $T_{e^{+}-e^{-}}$ the 
Reynolds number drops below 1. This is the result of the sudden drop in the number of charged particles that leads to a rather long mean free path for 
the photons. }.

\begin{figure}
\centerline{\epsfxsize = 9cm  \epsffile{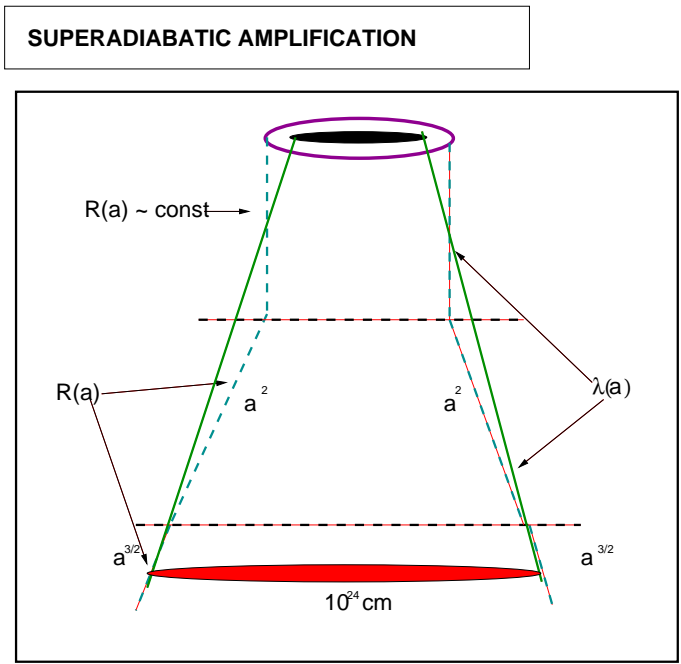}}
\vskip 3mm
\caption[a]{Evolution of the correlation scale if magnetic fields would be produced 
 by superadiabatic amplification during a conventional inflationary phase. The dashed 
 vertical lines denote, in the present figure, the evolution of the Hubble 
 radius while the full line denotes the evolution 
 of the correlation scale (typically selected to smaller than the Hubble radius during inflation).}
\label{FIG2} 
\end{figure}

In the context of (conventional or unconventional) 
inflationary models, on the contrary, the correlation 
 scale of the produced magnetic fields may be rather high. 
The physical picture is that during the inflationary phase 
 the quantum fluctuations of the hypercharge field are amplified either 
 through some coupling to the geometry or through the coupling 
 to some other spectator field. Since during inflation the Hubble 
 radius is roughly constant (see Fig. \ref{FIG2}), the correlation scale evolves much faster than the Hubble radius itself and, therefore, large scale 
 magnetic domains can naturally be obtained. Notice that, in Fig. \ref{FIG2}
 the (vertical) dashed lines illustrate the evolution of the Hubble radius 
 (that is roughly constant during inflation) while the full line denotes 
 the evolution of the correlation scale. Furthermore, the horizontal (dashed) lines 
 mark, from top to bottom, the end of the inflationary phase and the onset of the 
 matter-dominated phase.
 This phenomenon
 can be understood as  the gauge counterpart of the super-adiabatic 
 amplification of the scalar and tensor modes of the geometry.
The main problem, in such a framework, is to 
get large amplitudes  for scale of the order of the ${\rm Mpc}$ at the 
onset of gravitational collapse. Models where the gauge couplings 
are effectively dynamical (breaking, consequently, the Weyl invariance 
of the evolution equations of Abelian gauge modes) 
may provide rather intense magnetic fields.

The two extreme possibilities 
mentioned above may be sometimes combined.
 For instance, it can happen that 
magnetic fields are produced by super-adiabatic amplification of vacuum 
fluctuations during an inflationary stage of expansion. After 
exiting the horizon, the  gauge modes  will reenter at different moments 
all along the radiation and matter dominated epochs.
The spectrum of the primordial gauge fields after reentry 
 will not only be determined by the amplification mechanism but also on the plasma effects. As soon as the magnetic inhomogeneities reenter, some 
other physical process, taking place inside the Hubble radius, may be triggered by the presence of large scale magnetic fields. An example, in this 
context, is the production of topologically non-trivial configurations 
of the hypercharge field (hypermagnetic knots) from a stochastic 
background of hypercharge fields with vanishing helicity.

Since magnetic fields are produced over different length scales, it seems 
plausible to discuss their effect on the anisotropies of the 
Cosmic Microwave Background radiation (CMB). 
The finite values of the thermal and magnetic diffusivity scales 
forbids the presence of magnetic fields over length-scales that are 
very small in comparison with the size of the Hubble radius at a given epoch.
On the other hand, dissipation 
is not effective in erasing the magnetic energy density over very large 
scales comparable with the Hubble radius itself.

Different topics are relevant for the interplay 
between large-scale magnetic fields and CMB physics. 
A simplified list of the main effects involving, simultaneously, 
 large-scale magnetic fields and CMB physics must include:
\begin{itemize}
\item{} possible distortions of the CMB blackbody spectrum;
\item{} possible induction of primary anisotropies;
\item{} possible effects on the CMB polarization.
\end{itemize}
To this list it is appropriate to add another important topic that has to do 
with the subtraction of foregrounds. In fact, to disentangle 
the signal of the CMB anisotropies one has usually to subtract 
the emission of the galaxy. The synchrotron emission is indeed one of the foregrounds to be subtracted. 

To estimate the effects of large-scale magnetic fields on CMB 
physics various approximations have been used through the years.
Again, for sake of simplicity, it is useful to divide them into two 
broad classes:
\begin{itemize}
\item{} uniform field approximation;
\item{} inhomogeneous field approximation.
\end{itemize}
In the uniform field approximation, the magnetic field is taken to be 
uniform and pointing along a specific direction in the sky. In 
this approximation (pioneered by Zeldovich in the sixties) the 
magnetic field slightly breaks the spatial isotropy of the background 
geometry. The 
break in the spatial isotropy can be easily 
connected with a difference in the propagation of electromagnetic signals 
along different spatial directions. This phenomenon is ultimately 
responsible, in this context, for the anisotropy of the CMB.

In a complementary perspective, fully inhomogeneous
large-scale magnetic fields do not break the spatial isotropy 
of the background geometry. This happens, for instance,
 when the magnetic field is fully inhomogeneous 
and characterized by its two-point function. 
In the uniform field approximation 
the resulting evolution equations are rather tractable and 
semi-analytical estimates of various effects can be swiftly 
obtained.  It is, however, problematic to imagine 
that a uniform magnetic field (oriented along a specific spatial
direction)  may be generated in the early Universe. 
The presence of a uniform field could be viewed as a remnant 
of the initial conditions of the evolution of the Universe. 

Conversely, the fully inhomogenous approach is more realistic but also 
harder to implement analytically.
An intermediate approach is to discuss the uniform field case 
supplemented by the appropriate fluctuations. In this approach 
the parameters describing the model are given by the 
uniform component of the magnetic field and by the two-point 
function of the fluctuations.
 
An important point to be borne in mind when 
discussing the effects of large-scale magnetic fields on CMB 
anisotropies concerns the suitable set of equations used 
to model the plasma effects. For the purposes of this 
paper we will confine ourselves to the two most 
practical sets of equations normally employed in this 
type of studies, namely 
\begin{itemize}
\item{} a set of two-fluid plasma equations in curved space;
\item{} an effective one-fluid (magnetohydrodynamical) description.
\end{itemize}
These two systems of equations 
are not equivalent for all the frequencies of the spectrum 
of plasma excitations. In particular, if we ought to address 
phenomena involving the propagation of electromagnetic disturbances 
in the plasma (as in the case of Faraday) rotation, the two fluid 
description (or even a kinetic Vlasov-Landau description) have to be employed. Magnetohydrodynamics (MHD) is a good effective description 
for sufficiently small frequencies in the spectrum of plasma excitations. 

The plan of the present review is the following. In Section 2 the uniform 
field approximation will be reviewed. In Section 3 a two-fluid plasma 
description will be introduced. As an application, the Faraday effect 
in the uniform field approximation will be discussed.
In Section 4, after discussing a one-fluid plasma description,
 fully inhomogeneous magnetic fields will be analyzed. 
 Section 5 contains an introduction to the evolution of metric 
 fluctuations in the presence of large-scale magnetic fields. In this 
 context, the problem of initial conditions of CMB anisotropies 
 will be reviewed in the presence of large-scale magnetic fields.
 The final part of Section 5 contains a summary of the numerical results 
 obtained so far  in the context of fully inhomogeneous magnetic fields.
 A brief collection of concluding remarks and future perspectives 
 is collected in Section 6.

\renewcommand{\theequation}{2.\arabic{equation}}
\section{Uniform field approximation}
\setcounter{equation}{0}
The uniform field approximation has a long history that could be traced back to 
the work of Zeldovich \cite{zelm1} (see also \cite{zelm2,kal1}). 
Before the formulation of inflationary models the problem of the  the initial conditions of Friedmann-Robertson-Walker models was debated. 
Among different options 
(early dominance of viscous effects, anisotropy in the expansion, etc.) a particularly interesting possibility was represented by a uniform background
magnetic field oriented along a specific spatial direction.
Indeed, close to a big-bang 
singularity the spatial gradients are sub-leading but the model can well 
by anisotropic \cite{lif1,lif2,BK1}. 
In other words, close to the big bang,
 the homogeneous (but anisotropic) metric can 
fall in one of the Bianchi classes \cite{RS} and a uniform magnetic field 
can fit into this scheme. During an inflationary phase the anisotropy 
decays exponentially and, therefore, we have to imagine that the 
(uniform) field considered in this section was created, for instance, 
close to the end of inflation.
Since the work of Zeldovich \cite{zelm1,zelm2}, the uniform 
field approximation has been
exploited in various frameworks 
(see Refs. \cite{unif1,lim1,giofully,bama,giohom1,bronikov} and references 
therein for a tentative list of theoretical studies).

Consider then, as an illustration, 
the case when the magnetic field is oriented along the $\hat{x}$ axis. 
Suppose also, for sake of simplicity, that the Universe is filled by a 
perfect barotropic fluid and that the spatial curvature vanishes. 
In this case the line element must account for the possibility of a 
different expansion rate along the $\hat{x}$ axis or along the $(\hat{y},
\hat{z})$ plane and may have the form
\begin{equation}
ds^2 = dt^2 - a^2(t) dx^2 - b^2(t) ( dy^2 + dz^2).
\label{anism}
\end{equation}
The metric (\ref{anism}) corresponds to a particular 
case of a Bianchi-type I metric.  Uniform magnetic fields
can be introduced also in other Bianchi metrics 
\cite{unif1,giofully,bama,giohom1}.
The anisotropy in the expansion and the total expansion 
rate are, in this case,
\begin{equation}
{\cal A} = \frac{ H - F}{\Theta},\qquad \Theta = \frac{ H + 2 F}{ 3},
\label{def1}
\end{equation}
where $H = \dot{a}/a$ and $F= \dot{b}/b$ are the expansion 
rates and the overdot denotes a derivation with respect to the 
cosmic time coordinate $t$. 
\begin{figure}
\centerline{\epsfxsize = 9cm  \epsffile{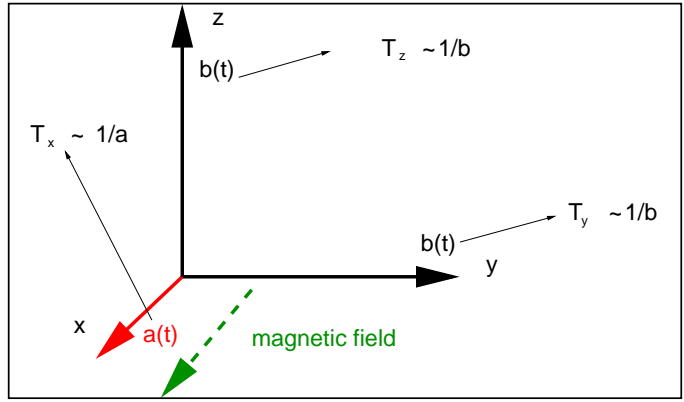}}
\vskip 3mm
\caption[a]{Uniform field approximation. A simplified geometrical 
configuration where the magnetic field is directed along the $\hat{x}$ axis.}
\label{FIG3} 
\end{figure}

In the model defined by Eq. (\ref{anism}) the propagation of an electromagnetic signal will be different along the $\hat{x}$ and $\hat{y}$ 
axes (see Fig. \ref{FIG3}). 
Therefore, the temperature anisotropy will depend both on $\Theta$ and 
${\cal A}$:
\begin{equation}
\frac{\Delta T}{T} = \frac{T_{x} - T_{y}}{T_{x}} 
\simeq - \int {\cal A} \,\, \Theta d t, 
\label{anis1}
\end{equation}
where we used that $T_{x} = T_{0}/a$ and $T_{y} = T_{0}/b$ together 
with the identities $a^{-1} = e^{- \int H\,d t}$ and $b^{-1} =e^{- \int F\,d t}$.

The quantities appearing in Eq. (\ref{anis1}) can be estimated by 
studying the evolution of the anisotropic expansion and of the total expansion
rate along the different stages of the life of the Universe. For this 
purpose it is practical to phrase Einstein equations directly in terms of
${\cal A}$ and $\Theta$.  The procedure is rather simple 
since it suffices to write the extrinsic curvature in terms of $H$ and $F$.
Then, by recalling the definitions of Eq. (\ref{def1}), the full set of Einstein 
equations and of the continuity equations can be written, respectively, as
\begin{eqnarray}
&& \dot{\Theta} + 3 \Theta^2  = \frac{8\pi G}{3} \rho \biggl[ \frac{2 - 3 \epsilon}{2} + q \biggr],
\label{TH1}\\
&& \dot{\cal A} \Theta + {\cal A} (\dot{\Theta} + 3 \Theta^2) = 
- 16\pi G \rho q,
\label{TH2}\\
&& \dot{\rho} + \Theta( 4 + 3 \epsilon) \rho =0,
\label{TH3}\\
&& \dot{q} = 3 q \Theta \epsilon + \frac{4}{3} q \Theta {\cal A},
\label{TH4}
\end{eqnarray}
where the following conventions have been adopted
\begin{equation}
w = \epsilon + \frac{1}{3},\qquad q = \frac{{\rho}_{\rm B}}{\rho}.
\label{conv1}
\end{equation}
In Eq. (\ref{conv1}) $w$ is the barotropic index of the fluid 
background and 
\begin{equation}
\rho_{\rm B} = \frac{{\cal B}^2}{8\pi},\qquad \dot{\rho}_{\rm B} + 4 F \rho_{\rm B} =0.
\label{magnevol}
\end{equation}
In the case $\epsilon =0$ the barotropic index is exactly the one 
of radiation and Eqs. (\ref{TH1})--(\ref{TH4}) admit 
a solution where $ \dot{{\cal A} }\simeq 0$ and $\Theta \simeq (2 t)^{-1}$. 
Inserting then Eq. (\ref{TH2})  into Eq. 
(\ref{TH4})  (and recalling that $\dot{\Theta} = - 2 \Theta^2$) 
it is easy to obtain the evolution of $q(t)$, i.e.
\begin{equation}
q(t) = \frac{ q_1}{ 1 + 4\,q_{1}\ln{\biggl(\frac{t}{t_1}\biggr)} }, \qquad t \geq t_{1}.
\label{qevol}
\end{equation}
This means that in the case of a radiation background the anisotropy 
in the expansion is conserved and the ratio between radiation and 
magnetic energy densities is roughly constant (up to logarithmic corrections).
For generalization of this analysis to more complicated situations 
involving different equations of state or various fluids and fields 
see, for instance, \cite{unif1,giofully,bronikov}.

The property expressed by Eq. (\ref{qevol}) is a consequence 
of the fact that both energy-momentum tensors (i.e. the one 
of the background fluid and the one of the magnetic field) are traceless. Therefore, the evolution of $q(t)$ will be different if the background fluid is dominated by dusty sources. 
In the case of dusty matter Eq. (\ref{conv1}) implies that 
$\epsilon = -1/3$. Therefore, an approximate solution of the system in this 
case is given by 
\begin{equation}
\Theta(t) \simeq \frac{2}{3\, t},\qquad {\cal A}(t) \simeq - 12 q_{\rm eq} \biggl(\frac{a_{\rm eq}}{a}\biggr),\qquad q(t) = q_{\rm eq} \biggl(\frac{a_{\rm eq}}{a}\biggr).
\end{equation}
By following the evolution of the system through equality up to decoupling 
the value of the anisotropy can then be roughly calculated to be 
\begin{equation}
\frac{\Delta T}{T} \simeq \frac{{\cal B}^2}{16\pi \rho_{\gamma}} z_{\rm dec}.
\label{anisB}
\end{equation}
By requiring that the temperature anisotropy is smaller than $10^{-5}$ we get 
in the case of $\Omega_{\rm tot}= 1$, 
\begin{equation}
{\cal B}_{0} \leq 2 \times 10^{-9} \,\,\,\, {\rm Gauss},
\label{B1}
\end{equation}
where ${\cal B}_{0}$ refers to the value of the field at the present time.
The logic leading to the limit reported in Eq. (\ref{B1}) has been 
followed, for instance, in Ref. \cite{lim1}.

A complementary way of describing the effect of a uniform magnetic field 
on the temperature anisotropies relies on the observation that, thanks 
to the breaking of spatial isotropy,  the angular power spectrum 
will not depend only on the polar angle but also 
on the azimuthal angle. 
In fact, if spatial isotropy is  {\em unbroken}, the angular power spectrum 
is customarily defined as 
\begin{equation}
C(\vartheta) = \langle \Delta_{\rm I}(\hat{n}_{1},\tau_{0})  \Delta_{\rm I}(\hat{n}_{2},\tau_{0}) \rangle,
\label{PS}
\end{equation}
where $ \Delta_{\rm I}$ is the brightness perturbation in the intensity 
of the radiation field whose expansion in terms of spherical harmonics 
is 
\begin{equation}
\Delta_{\rm I}(\hat{n},\tau_{0} ) = \sum_{\ell=0}^{\infty}
 \sum_{m=-\ell}^{\ell} a_{\ell m} Y_{\ell m}(\hat{n}).
\label{ANIS}
\end{equation}
If the background space is isotropic, the ensemble average 
of $a_{\ell\,m}$ only depends upon $\ell$, not upon $m$, i.e. 
\begin{equation}
\langle a_{\ell m}a^{\ast}_{\ell' m'} \rangle = C_{\ell} \delta_{\ell\ell'} \delta_{m m'}.
\label{AVALM}
\end{equation}
Therefore, inserting Eq. (\ref{ANIS}) into Eq. (\ref{PS}), the addition theorem of spherical harmonics, i.e. 
\begin{equation}
P_{\ell }(\hat{n}_1\cdot \hat{n}_2)  = \frac{4\pi}{2\ell + 1}\sum_{m=-\ell}^{\ell} Y_{\ell m}^{\ast}(\hat{n}_{1}) Y_{\ell m} (\hat{n}_{2}),
\label{ADD}
\end{equation}
leads to the usual form of the angular power spectrum:
\begin{equation}
C(\vartheta) =  \langle \Delta_{\rm I}(\hat{n}_{1},\tau_{0})  \Delta_{\rm I}(\hat{n}_{2},\tau_{0}) \rangle \equiv
\frac{1}{4\pi} \sum_{\ell} (2 \ell + 1) C_{\ell } P_{\ell}(\vartheta).
\label{Cth}
\end{equation}
where $\vartheta =\hat{n}_{1} \cdot \hat{n}_{2}$.

If spatial isotropy is slightly broken 
(as in the case when a uniform magnetic field 
is present), Eq. (\ref{AVALM}) is not justified since the ensemble average 
of the $a_{\ell\,m}$ should also depend upon $m$ and not only upon $\ell$.
Indeed, it can be shown that when a uniform magnetic field is present, 
the average of temperature fluctuations in two directions of the sky 
can be expressed as 
\begin{eqnarray}
&&\langle \Delta_{\rm I}(\hat{n}_{1},\tau_{0}) \Delta(\hat{n}_{2},\tau_{0}) \rangle =
\sum_{\ell,\,m} C_{\ell}(m) Y_{\ell,\, m}(\hat{n}_1) Y^{\ast}_{\ell,\,m}(\hat{n}_2) 
\nonumber\\
&&+ \sum_{\ell,\,m} D_{\ell}(m)[ Y_{\ell + 1,\,m}(\hat{n}_{1}) Y^{\ast}_{\ell -1,\,m}(\hat{n}_{2}) + Y_{\ell-1,\,m}(\hat{n}_{1}) Y^{\ast}_{\ell + 1,\,m}(\hat{n}_{2})],
\label{Cthph}
\end{eqnarray}
where both $C_{\ell}(m)$ and $D_{\ell}(m)$ depend upon the (uniform) magnetic field
intensity. From Eq. (\ref{Cthph}) it is apparent 
that the presence of a magnetic field 
with uniform component induces off-diagonal correlations in multipole 
space between a given $\ell$ and  multipoles $\ell \pm 2$. The breaking 
of spatial isotropy in the CMB maps can be constrained, for 
instance, from WMAP data . 

This analysis has been recently performed in \cite{chen} (see also \cite{yates}). The constraints 
on the off-diagonal correlations in multipole space imply a bound 
on the (present) uniform component of the magnetic field intensity 
\begin{equation}
{\cal B}_{0} \leq 1.7\times 10^{-9} \,\, {\rm G}.
\label{B2}
\end{equation}
It should be mentioned that the bound of Eq. (\ref{B2}) has been 
derived by taking into account also the fluctuations of the magnetic field
around the uniform configuration. The presence of the fluctuations of a 
uniform configuration introduces extra parameters (like the spectral 
index of the fluctuations) that make the allowed region in the parameter 
space effectively two-dimensional.
Various studies assume a (uniform) background field and consider 
the possible fluctuations of such a configuration (see for instance 
\cite{b1,b2,b3} and references therein). In particular, in Ref. \cite{b2} 
the magnetic field has been parametrized in terms of a uniform 
component directed along a specific axis and a tangled component. The 
Alf\'en-waves modes associated with this configuration may induce a small 
rotational perturbation in the last scattering surface.

In \cite{rg} this analysis has been performed in a non-relativistic 
approximation where the scalar perturbations of the geometry obey linearized Newtonian equations of motion. The magnetic field has been 
assumed to be uniform (but supplemented with the appropriate fluctuations). In this framework, it has been argued that the 
effect of the presence of the (uniform)  magnetic can be phrased into an effective renormalization of the speed of sound of the baryons (see also 
\cite{rgrep}).

\renewcommand{\theequation}{3.\arabic{equation}}
\section{Two-fluid plasma description and CMB polarization}
\setcounter{equation}{0}
Right before decoupling the value of the Debye screening length is much larger than the electron and photon mean free paths. In fact, the Debye length scale can be written as 
\begin{equation}
\lambda_{\rm D} = \sqrt{ \frac{T_{\rm ei}}{8\pi e^2 n_{0}}} \simeq 
 10\,\, \biggl(\frac{n_{0}}{10^{3}\,\,\, {\rm cm}^{-3}}\biggl)^{-1/2} 
\biggl( \frac{T_{\rm ei}}{0.3 \,\,\,{\rm eV}}\biggr)^{1/2} \,\, {\rm cm}.
\label{debye}
\end{equation}
where $n_{0} \simeq n_{\rm e} \simeq n_{\rm i}$ is the mean electron-ion density 
and $T_{\rm ei}$ is the electron-ion temperature. For  length scales 
larger than $\lambda_{\rm D}$, the plasma is globally neutral.
For typical values of the parameters in Eq. (\ref{debye}), 
$\lambda_{\rm D}$  is of the order of $10$ cm around 
recombination while the electron and photon mean free paths 
are much larger, i.e. 
\begin{equation}
\ell_{\rm e} \simeq 5.7 \times 10^{7} {\rm cm},\qquad \ell_{\gamma} \simeq 10^{4} (1 + z)^{-2} (\Omega_{\rm b} h_{0}^2)^{-1} \,\,\,{\rm Mpc}, 
\end{equation}
where $z = a_{0}/a -1$ is the redshift, $\Omega_{\rm b} $ the baryon fraction of the 
critical energy density and $h_{0} $ the indetermination in the Hubble 
parameter. 

The other set of quantities relevant for the determination of the appropriate system of equations is given by the plasma and Larmor frequencies of electrons and ions, i.e.
\begin{eqnarray}
&&  \omega_{\rm pe} \simeq
2\,\, \biggl( \frac{n_{e}}{10^{3} {\rm cm }^{-3}} \biggr)^{1/2} \,\, {\rm MHz},
\,\,\,\,\,\,\, \omega_{\rm Be}  \simeq 
18.08 \biggl(\frac{{\cal B}}{ 10^{-3}\,\,\, {\rm G}}\biggr)\,\,\,{\rm kHz},
\label{ef}\\
&&  \omega_{\rm pi} 
\simeq 40\,\, \biggl( \frac{n_{e}}{10^{3} {\rm cm }^{-3}} 
\biggr)^{1/2} \,\, {\rm kHz},
\,\,\,\,\,\,\, \omega_{\rm Bi}  \simeq  9.66  
\biggl(\frac{{\cal B}}{10^{-3}\,\,\, {\rm G}}\biggr)\,\,\,{\rm Hz}.
\label{if}
\end{eqnarray}
Note that the typical value of ${\cal B}$ around decoupling
 reported in Eqs. (\ref{ef}) and (\ref{if}) 
corresponds, today, to a magnetic field of the order of $10^{-9}$-$10^{-10}$ G.

If we are now interested in the propagation of electromagnetic 
disturbances with frequencies larger than the plasma and Larmor 
frequencies, then it is clear that a two-fluid plasma description should 
 be adopted.
For some classes of problems it will also be useful to employ directly 
a kinetic (Vlasov-Landau \cite{vl1,vl2,ll1}) description.
 If, on the contrary, we would be interested 
in the spectrum of plasma excitations for $ \omega < \omega_{\rm p}$, then 
effective (one-fluid) descriptions (like MHD \cite{biskamp}) can be usefully employed. This second regime will be addressed in Section 4.

The two-fluid plasma equations in curved space-time will now be discussed 
and treated in analogy with the well-known flat-space case \cite{boyd}.
Let us consider, for simplicity a conformally flat geometry of 
Friedmann-Robertson-Walker (FRW) type:
\begin{equation}
ds^2 = a^2(\tau) [ d\tau^2 - d \vec{x}^2].
\end{equation}
The following two equations account for the continuity of 
electrons and ions charge densities
\begin{eqnarray}
&& n_{\rm e}' + 3 w_{\rm e} {\cal H} n_{\rm e} 
+ ( w_{\rm e} + 1 ) \vec{\nabla}\cdot( n_{\rm e} \vec{v}_{\rm e} ) =0,
\label{ne}\\
&& n_{\rm i}' + 3 w_{\rm i} {\cal H} n_{\rm i} 
+ ( w_{\rm i} + 1 ) \vec{\nabla}\cdot( n_{\rm i} \vec{v}_{\rm i} ) =0,
\label{ni}
\end{eqnarray}
where the prime denotes a derivation 
with respect to the conformal time coordinate $\tau$ while the 
other quantities appearing in Eqs. (\ref{ne}) and (\ref{ni})  are
\begin{equation}
{\cal H} = a'/a, \qquad n_{\rm e} = a^{3} \tilde{n}_{\rm e}, \qquad 
n_{\rm i} = a^{3} \tilde{n}_{\rm i}.
\end{equation}
In Eqs. (\ref{ne}) and (\ref{ni}) 
$w_{\rm e}$ and $w_{\rm i}$ are the  barotropic indices 
for the electron and ion fluids.
Both electrons and ions are non-relativistic (for the ion mass, in the numerical
estimates, we take $m_{\rm i} \simeq m_{\rm p}$ where 
$m_{\rm p}$ denotes the proton mass). 
Hence the barotropic indices $w_{\rm e}$ and $w_{\rm i}$ will 
be close to zero. In fact, the energy and pressure densities 
of an ideal electronic gas are given by 
\begin{equation}
\rho_{\rm e} = n_{\rm e} \biggl( m_{\rm e} 
+ \frac{3}{2} T_{\rm e}\biggr),\,\,\,\,\,\,\,\,\,\,
p_{\rm e} = n_{\rm e} T_{\rm e},
\end{equation}
and since $ w_{\rm e,i} = T_{\rm e,i}/m_{\rm e,i}$,  we will also have 
$ w_{\rm e,i} \ll 1$ as far as $ T_{\rm e,i} \ll m_{\rm ,ie}$. 
Recalling that $\rho_{\rm e} = m_{\rm e} \, n_{\rm e} $ and $\rho_{\rm i} = 
m_{\rm i} n_{\rm i}$, the momentum conservation equations 
are
\begin{eqnarray}
&& \rho_{\rm e} [ \vec{v}_{\rm e}' + {\cal H} \vec{v}_{\rm e} 
+ (\vec{v}_{\rm e}\cdot\vec{\nabla})\vec{v}_{\rm e} ] = - n_{\rm e}
e \biggl( \vec{E} + \frac{\vec{v}_{\rm e}}{c} \times \vec{B}\biggr),
\label{ve}\\
&&   \rho_{\rm i} [ \vec{v}_{\rm i}' + {\cal H} \vec{v}_{\rm i} 
+ (\vec{v}_{\rm i}\cdot\vec{\nabla})\vec{v}_{\rm i} ] =  n_{\rm i}
e \biggl( \vec{E} + \frac{\vec{v}_{\rm i}}{c} \times \vec{B}\biggr),
\label{vi}
\end{eqnarray}
where $\vec{E} = a^2 \vec{{\cal E}}$ and $\vec{B} = a^2 \vec{{\cal B}}$ 
are the conformally rescaled electromagnetic fields obeying 
the following set of equations:
\begin{eqnarray}
&& \vec{\nabla} \cdot \vec{E} = 4 \pi e ( n_{\rm i} - n_{\rm e} ),
\label{div1}\\
&&\vec{\nabla} \cdot \vec{B} =0,\,\,\,\,\,\,\,\,\,\,\,\vec{\nabla} \times \vec{E} = 
- \frac{1}{c} \vec{B}',
\label{bianchi1}\\
&& \vec{\nabla} \times  \vec{B} = \frac{1}{c} \vec{E}' 
 + \frac{ 4 \pi e}{c} ( n_{\rm i} \vec{v}_{\rm i} - n_{\rm e} \vec{v}_{e} ).
\label{current}
\end{eqnarray}
The dispersion relations relevant for the propagation of electromagnetic waves 
with $\omega > \omega_{\rm p}$ can be found, after some algebra, in the case 
of a uniform magnetic field directed, for instance, along the $\hat{z}$ axis\footnote{In principle we should allow a different expansion rate 
along the $(\hat{x},\hat{y})$ plane and along the $\hat{z}$ axis, as discussed 
in the previous Section (see, for instance, Eq. \ref{anism}). 
This is not strictly necessary for 
the derivation of the dispersion relations. While the anisotropic 
expansion caused by a uniform magnetic field affects  the primary 
anisotropies, the dispersion relations are weakly affected by the 
curved space corrections. This statement will be explicitly verified in our case.}.

 Equations (\ref{ne}) and (\ref{ni}) can be first linearized  around a globally neutral configuration, i.e.
 \begin{equation}
n_{\rm e}(\tau, \vec{x}) = n_{0} + \delta n_{\rm e}(\tau,\vec{x}),
\qquad
 n_{\rm i}(\tau, \vec{x}) = n_{0} + \delta n_{\rm i}( \tau,\vec{x}),
\end{equation}
where $n_{0}$ is the common electron and ion density.
The remaining equations can be also linearized around the uniformly 
magnetized configuration:
\begin{eqnarray}
&& \vec{B}(\eta,\vec{x}) = \vec{B}_{0} + \delta \vec{B}(\eta,\vec{x}),
\nonumber\\
&& \vec{v}_{\rm e,\,i}(\eta,\vec{x}) = 
\delta \vec{v}_{\rm e,\,i} (\eta,\vec{x}),\,\,\,\,\,\,\,\,\,\,\,\,\,\,\,\,\,\,\,\,\,\
\vec{E}(\eta,\vec{x})= \delta \vec{E}(\eta,\vec{x}).
\label{fluct}
\end{eqnarray}
The linearized equations around the uniformly magnetized background 
can be solved \cite{birefringence}.
The dielectric tensor can then be obtained and it 
has the following general form \cite{birefringence}:
\begin{equation}
\overline{\epsilon}(\omega,\alpha) = \left(\matrix{\epsilon_{1}(\omega,\alpha)
& i\epsilon_{2}(\omega,\alpha) & 0&\cr
-i \epsilon_{2}(\omega,\alpha) & \epsilon_{1}(\omega,\alpha) &0&\cr
0&0&\epsilon_{\parallel}(\omega,\alpha) }\right),
\label{epstens}
\end{equation}
where 
\begin{eqnarray}
&&\epsilon_{\parallel}(\omega,\alpha) = 1 - \frac{ \omega_{\rm p,\,i}^2}{\omega^2 (1 +\alpha)} - \frac{ \omega_{\rm p,\,e}^2}{\omega^2 (1 +\alpha)}.
\label{epspar}\\
&& \epsilon_{1}(\omega,\alpha) = 1 - \frac{\omega^2_{\rm p\, i} (\alpha + 1) }{\omega^2 (\alpha+ 1)^2 - \omega_{\rm B\, i}^2} -
\frac{\omega^2_{\rm p\, e} (\alpha + 1) }{\omega^2 (\alpha+ 1)^2 - \omega_{\rm B\, e}^2},
\label{eps1}\\
&& \epsilon_{2}(\omega,\alpha) = \frac{\omega_{\rm B\, e}}{\omega} \frac{\omega^2_{\rm p\, e } }{\omega^2 (\alpha+ 1)^2 - \omega^2_{\rm B\, e}} 
-  \frac{\omega_{\rm B\, i}}{\omega} \frac{\omega^2_{\rm p\, i } }{\omega^2 (\alpha+ 1)^2 - \omega^2_{\rm B\, i}} .
\label{eps2}
\end{eqnarray}
The quantity $\epsilon_{\parallel}(\omega,\alpha)$ denotes the dielectric constant 
parallel to the direction of the uniform component of the magnetic field 
intensity while $\epsilon_{1}(\omega,\alpha)$ and 
$\epsilon_{2}(\omega,\alpha)$ denote the dielectric constants 
determined by the motion of charged particles orthogonal to the direction of 
the uniform magnetic field. If the uniform magnetic field vanishes, we can obtain the dielectric tensor of the cold plasma by formally taking the 
limit
$\epsilon_{2}(\omega,\alpha) \to 0$ and $\epsilon_{1}(\omega,\alpha) \to \epsilon_{\parallel}(\omega,\alpha)$. 
It is then clear that in the absence of magnetic field 
the only scales entering the dispersion relations are the plasma frequencies of 
electrons and ions.  The quantity $\alpha$ appearing in the 
dielectric tensor is defined as $ \alpha = i {\cal H}/\omega$ and 
is the lowest order correction arising in curved space. This 
correction is negligible for the present purposes since it is of the order of the 
ratio between the Hubble radius at the decoupling and the inverse of the 
frequency of propagation. Recalling that $ H_{\rm dec} \simeq 5.6 \times 
10^{12} 
(\Omega_{0} h_{0}^2)^{-1/2} \,\, {\rm sec}$ and that $ \omega \gg 100\,\,
{\rm MHz}$, we  have that $|\alpha| < 10^{-22}$.

To derive the dispersion relations it is appropriate to fix the coordinate system:
for instance we can set  $k_{x}=0$ and 
$k_{y} = k \sin{\theta}$, $k_{z} = k \cos{\theta}$ with the magnetic field
oriented along the $\hat{z}$ direction. 
Introducing the refraction index $n$ and recalling that $k = \omega/v = n \omega/c$ (where $v$ is the phase velocity)
 the dispersion relations are determined from the solution of the following equation:
\begin{eqnarray}
&&s^2(\theta) \biggl( \frac{1}{\epsilon_{\parallel}} - \frac{1}{n^2} 
\biggr) \biggl[ \frac{1}{n^2} - \frac{1}{2} \biggl( \frac{1}{\epsilon_{\rm L}}
+ \frac{1}{\epsilon_{\rm R}}\biggr) \biggr] -
c^2(\theta)\biggl( \frac{1}{n^2} - \frac{1}{\epsilon_{\rm L}}
\biggr)\biggl( \frac{1}{n^2} - \frac{1}{\epsilon_{\rm R}}\biggr) =0,
\label{AH}
\end{eqnarray}
where $c(\theta) = \cos{\theta}$ and $s(\theta) = \sin{\theta}$. In Eq. (\ref{AH})
the following linear combinations have been also introduced: 
\begin{eqnarray}
&& \epsilon_{\rm R} = \epsilon_{1} + \epsilon_{2} = 1 - \frac{ \omega^2_{\rm p\,i}}{\omega[ \omega (\alpha + 1) - \omega_{\rm B\,i}]} 
-\frac{ \omega^2_{\rm p\,e}}{\omega[ \omega (\alpha + 1) 
+ \omega_{\rm B\,e}]},
\label{epsR}\\
&& \epsilon_{\rm L} = \epsilon_{1} - \epsilon_{2} = 1 -  \frac{ \omega^2_{\rm p\,e}}{\omega[ \omega (\alpha + 1) -\omega_{\rm B\,e}]}
- \frac{ \omega^2_{\rm p\,i}}{\omega[ \omega (\alpha + 1) 
+ \omega_{\rm B\,i}]}.
\label{epsL}
\end{eqnarray}
The quantity $\epsilon_{\rm R}$ and $\epsilon_{\rm L}$ 
are the dielectric constants appropriate for the propagation of 
the two (left and right) circular polarizations. The dispersion relations
for the propagation of electromagnetic disturbances parallel (i.e. $\theta =0$) and orthogonal (i.e. $\theta =\pi/2$) to the uniform component 
of the magnetic field can then be obtained from Eq. (\ref{AH}):
\begin{eqnarray} 
&& (n^2 - \epsilon_{\rm R})(n^2 -\epsilon_{\rm L})=0,\,\,\,\,\,\,
\,\,\,\,\,\,\,\,\,\,\,\,\,\,\,\,\,\,\,\,\,\,\,\,\,\,\,\,\theta=0,
\label{std0}\\
&& (n^2 - \epsilon_{\parallel}) [ n^2 (\epsilon_{\rm L} + \epsilon_{\rm R})
- 2 \epsilon_{\rm L} \epsilon_{\rm R} ] =0,
\,\,\,\,\, \theta = \frac{\pi}{2}.
\label{stdpi2}
\end{eqnarray}
Equation (\ref{std0}) gives the usual dispersion relations for the two circular 
polarizations of the electromagnetic wave, i.e. $n^2 = \epsilon_{\rm R}$ 
and $n^2 = \epsilon_{\rm L}$, while Eq. (\ref{stdpi2}) 
gives those for the ``ordinary'' 
(i.e. $n^2 = \epsilon_{\parallel}$) and ``extraordinary'' (i.e. $n^2 = 
2 \epsilon_{\rm R} \epsilon_{\rm L}/(\epsilon_{\rm R} + \epsilon_{\rm L})$)
plasma waves.  

Consider now  an electromagnetic wave that is 
linearly polarized. A linearly polarized electromagnetic wave 
can be always seen as the sum of two 
circularly polarized waves. Now the left and right polarizations have 
different phase velocities and, therefore, after the waves 
have travelled a distance $\Delta z$ in a (cold) magnetized plasma the 
resulting phase difference will be 
\begin{equation}
\Delta\Phi = \frac{\omega}{2c} \biggl[ 
 \sqrt{\epsilon_{\rm R}} -\sqrt{\epsilon_{\rm L}} \biggr] \Delta z.
\label{FE}
\end{equation}
Equation (\ref{FE}) has been derived using two physical assumptions 
\begin{itemize} 
\item{} the plasma is cold, i.e. the temperature of the electrons 
does not enter the dispersion relations;
\item{} there are no other sources leading to a rotation of linearly 
polarized radiation.
\end{itemize}
Let us now see what happens by relaxing these two assumptions.
The adoption of a cold (as opposed to warm) plasma 
description means that the temperature 
of the electrons has been taken to be effectively zero. 
The effects arising from the finite temperature of the
 electrons do not modify the leading result obtained in the context of the 
cold plasma theory. The derivation of the dispersion relations 
in the case of a warm plasma can be performed, for instance, within a 
kinetic approach where it can be shown, following the same 
calculation discussed in the flat space case \cite{skilling}, that the first correction to the leading cold plasma calculation can be
 recast in an effective redefinition of the 
plasma frequency for the electrons, namely
$\omega_{\rm p\,e} \to \frac{4 \pi n_{0} e^2}{m_{e} \gamma}$
where $\gamma = (1 - \langle v^2\rangle)^{-1/2}$  and $\langle v^2 \rangle$ is the thermal average of the electron velocity. 

 Equation 
(\ref{FE}) has been derived under the assumption 
that the only source of birefringence is represented 
by a uniform magnetic field. This is certainly legitimate 
in a standard electromagnetic context. There could be however 
other sources of cosmological birefringence leading 
to a generalized expression of Faraday rotation.
In particular let us consider the case when, in the plasma, a dynamical 
pseudo-scalar field is present together with a magnetic field.  This 
situation may easily arise in the case of a quintessential background 
with pseudo-scalar coupling to electromagnetism described by an action 
\begin{equation} 
S_{\sigma} = \int d^{4} x \sqrt{-g} \biggl[ \frac{1}{2}g^{\alpha\beta} 
\partial_{\alpha}\sigma \partial_{\beta} \sigma - W(\sigma) + 
\frac{\beta}{4 M} F_{\alpha\beta} \tilde{F}^{\alpha\beta} \biggr],
\label{PNGB}
\end{equation}
where $F_{\alpha\beta}$ and $ \tilde{F}^{\alpha\beta}$ are the Maxwell 
field strength and its dual. The coupling of electromagnetism 
to the field $\sigma$ changes the form of Eqs. (\ref{div1}), (\ref{bianchi1}) 
and (\ref{current}). The resulting system of equations can be 
discussed in the presence both of a uniform magnetic field and of 
a cold (or warm) plasma. This calculation has been 
performed in Ref.  \cite{birefringence}.
While Eqs. (\ref{bianchi1}) do not change, the generalization
of  Eqs. (\ref{div1}) and (\ref{current}) becomes:
\begin{eqnarray}
&& \vec{\nabla} \cdot \vec{E} = 4 \pi e ( n_{\rm i} - n_{\rm e} ) 
+ \frac{\beta}{M} \vec{\nabla}\sigma \cdot \vec{B},
\label{div2}\\
&& \vec{\nabla} \times  \vec{B} = \frac{1}{c} \vec{E}' - \frac{\beta}{M} 
\biggl[\sigma' \vec{B} + \vec{\nabla}\sigma \times \vec{E} \biggr] 
 + \frac{ 4 \pi e}{c} ( n_{\rm i} \vec{v}_{\rm i} - n_{\rm e} \vec{v}_{e} ).
\label{current2}
\end{eqnarray}
To understand the rationale of the physical difference induced by the 
presence of $\sigma$ consider, indeed, the case when 
$ \nabla \sigma =0$ and the charge and current densities 
are vanishing. In this case, taking the curl of Eq. (\ref{current2}) 
and using the second of Eqs. (\ref{bianchi1}) we obtain the following 
equation
\begin{equation}
\vec{B}''  - c^2 \nabla^2 \vec{B} + \frac{\beta}{M} \,c\, \sigma'  \vec{\nabla}\times 
\vec{B} =0.
\label{simplified}
\end{equation}
Equation (\ref{simplified}) implies that the two polarizations of the 
magnetic field are mixed as a consequence of the time-dependence 
of $\sigma$.  If the plasma contribution is included, 
Eq. (\ref{AH}) is modified \cite{birefringence} . Consequently, the 
dispersion relations given in Eqs. (\ref{std0}) and (\ref{stdpi2}) 
are also modified:
\begin{eqnarray}
&& \biggl( n^2 -  \frac{\omega_{\sigma}}{\omega}n - \epsilon_{\rm R}\biggr) 
\biggl(n^2 + \frac{\omega_{\sigma}}{\omega} n - \epsilon_{\rm L}\biggr) =0,
\,\,\,\,\,\,\,\,\,\,\,\,\,\,\,\,\,\,\,\,\,\,\,\,\,\,\,\,\theta =0,
\label{nsd0}\\
&& n^4 - \biggl[ 2 \frac{\epsilon_{\rm L}\epsilon_{\rm R}}{\epsilon_{\rm L} 
+ \epsilon_{\rm R}} + \epsilon_{\parallel} 
+ \biggl(\frac{\omega_{\sigma}}{\omega}
\biggr)^2\biggr] n^2 + 2 \frac{\epsilon_{\parallel} \epsilon_{\rm L} 
\epsilon_{\rm R}}{ \epsilon_{\rm R} +\epsilon_{\rm L}} =0,
\,\,\,\,\,\theta = \frac{\pi}{2}.
\label{nsdpi2}
\end{eqnarray}
The presence of the quintessence field 
introduces a further frequency scale into the problem, namely 
$\omega_{\sigma} = c (\beta/M) \sigma' $. If the quintessence field 
dominates today (or between redshifts $0$ and $3$), 
$\sigma(t_0) \sim \Lambda^2/m$ 
where, typically, $ \Lambda\sim 10^{-3}\,\, {\rm eV}$, and 
$ m \sim 10^{-33} \,\,{\rm eV}$. The value of $\dot{\sigma}$ can be estimated, 
 and it is, today, $ \dot{\sigma}(t_0) \simeq m^2/H_0$.
Hence, recalling that prior to quintessential dominance, $\dot{\sigma}$ scales 
as $a^{-3}$, from the previous expressions  
\begin{equation}
\dot{\sigma}_{\rm dec} \simeq \biggl( \frac{H_{\rm dec}}{H_{0}}\biggr)^2 
\frac{m \Lambda^2}{H_{0}},\,\,\,\,\,\,\,\,\,\,\,
\omega_{\sigma} = \beta \biggl( \frac{M}{M_{\rm P}}\biggr) \times 10^{-6} \,\,
{\rm Hz},
\label{omegasigma}
\end{equation}
where the values of $\Lambda$ 
and $m$ are the ones discussed above with $M\simeq M_{\rm P}$. The value of 
$\beta$ is rather uncertain and a conservative limit from 
radio-astronomical analyses would imply $ \beta \laq 10^{-3}$  for
$0 \laq \,\, z\,\,\laq 1$ \cite{q2,q3}. 
Notice that the action (\ref{PNGB}) can also 
be relevant in a class of baryogenesis models \cite{q5,q6} (see also 
\cite{q7}).   

From Eq. (\ref{nsd0}) the generalized Faraday rotation experienced 
by the linearly polarized CMB travelling parallel to the magnetic 
field direction can be obtained as 
\begin{equation}
\Delta\Phi = \frac{\omega}{2c} \biggl[ \frac{\omega_{\sigma}}{\omega}
+ \sqrt{\frac{1}{2}  \biggl(\frac{\omega_{\sigma}}{\omega}\biggr)^2 
+ \epsilon_{\rm R}} -\sqrt{\frac{1}{2}  \biggl(\frac{\omega_{\sigma}}{\omega}\biggr)^2 + \epsilon_{\rm L}} \biggr] \Delta z.
\label{GENFAR}
\end{equation}
This expression includes, simultaneously, the birefringence 
produced by the evolution of the pseudo-scalar field $\sigma$ 
and the rotation induced by the finite value of the magnetic field 
intensity.  In Ref. \cite{breas} the effect of parity-violating 
interactions on CMB polarization  has been recently discussed
in the absence of a magnetic field.  A similar analysis was presented in 
\cite{lue} (see also \cite{lepora}). 
In the absence of magnetic field, 
Eq. (\ref{GENFAR}) leads to a $\Delta \Phi$ that 
 is frequency independent. On the contrary, if the contribution of 
 the pseudo-scalar field $\sigma$ is negligible, then $\Delta \Phi$ decreases 
 at high frequencies as $1/\omega^2$. In the intermediate 
 situation more complicated dependences of $\Delta\Phi$ upon 
 the frequency may be envisaged.   This aspect answers the second question 
 raised after Eq. (\ref{FE}): on top of the magnetic fields there are 
 different sources affecting $\Delta \Phi$. However, the characteristic 
 frequency dependence of the different signals may allow, at 
 least in principle, to disentangle different sources of birefringence.
 
\subsection{Magnetized Faraday effect at decoupling}

Let us now assume, for simplicity, that the only source 
for $\Delta\Phi$ is represented by a uniform magnetic field. Then,
recalling the hierarchy between the plasma, Larmor and propagation 
frequencies, i.e. Eqs. (\ref{ef}) and (\ref{if}),  Eq. (\ref{FE}) can be 
expanded for 
\begin{equation}
|\omega_{\rm p,\, e,\,i}/\omega| \ll 1,\qquad 
 |\omega_{\rm B,\, e,\,i}/\omega| \ll 1.
 \label{expan}
 \end{equation}
Thus, from Eq. (\ref{FE}), we will have: 
 \begin{equation}
 \Delta\Phi \simeq \frac{\omega}{2 c} \biggl\{ \sqrt{ 1 - 
 \biggl(\frac{\omega_{\rm p\,\,e}}{\omega}\biggr)^2\biggl[1 - \frac{\omega_{\rm B\,\,e}}{\omega}\biggr] }- \sqrt{ 1 - 
 \biggl(\frac{\omega_{\rm p\,\,e}}{\omega}\biggr)^2\biggl[1 + \frac{\omega_{\rm B\,\,e}}{\omega}\biggr]}  \biggr\} \,\,\Delta \,z,
 \label{inter1}
 \end{equation}
 where, by virtue of Eqs. (\ref{ef}) and (\ref{if}) the sub-leading contribution of the ions has been neglected. In some simplified treatments this approximation 
 is made from the very beginning by defining two generalized refraction 
 indices $n_{\rm R}$ and $n_{\rm L}$: 
 \begin{equation}
 n_{\rm R,\,\,L} = \sqrt{\epsilon_{\rm R,\,\,L}} = 1 - \frac{\omega_{\rm p,\,e}^2}{\omega( \omega \pm \omega_{\rm B,\,\,e})}
 \label{approx}
 \end{equation}
 ( the plus and minus signs correspond, respectively, to R and L\footnote{Recalling that $\omega = k \,c/n_{\rm R,\,L}$, by 
 compute $\partial\omega/\partial k$ we can easily get, from Eq. (\ref{approx}),
and approximate expression of the group velocity.})

A last expansion for 
 $|\omega_{\rm B,\,e}/\omega|\ll 1$ and for  $|\omega_{\rm p,\,e}/\omega|\ll 1$
 brings Eq. (\ref{inter1})  to the standard form, i.e. 
\begin{equation}
\Delta\Phi = \frac{1}{2 c} \biggl(\frac{\omega_{\rm p,\, e}}{\omega}\biggr)^2 \omega_{\rm B,\, e} \Delta z.
\label{FE2}
\end{equation}
From Eq. (\ref{FE2}), by choosing a generic propagation direction, we can also derive the Faraday rotation rate, i.e. 
\begin{equation}
\frac{d \Phi}{d t} = \frac{e^3\, x_{\rm e} n_0}{ 2\, \pi\,c\, m_{\rm e}^2 \, \nu^2} 
\,\,\,\vec{B}\cdot \hat{n}
\label{FE3}
\end{equation}
where the inonization fraction $x_{\rm e}$ has been introduced by replacing 
$n_{0}\to x_{\rm e } n_{0}$ and where we used, from Eq. (\ref{FE2}), that 
$dz = c\,dt$. In Eq. (\ref{FE3}), 
$\vec{B}\cdot \hat{n}$ denotes the projection of the 
(uniform) magnetic 
field intensity along the direction of propagation of the linearly polarized radiation with 
frequency $\nu = \omega/2\pi$; $t$ is, as usual, the cosmic time coordinate 
related to the conformal time coordinate $\tau$ as $ a(\tau) d\tau = dt $.

For some applications it is sometimes useful to write Eq. (\ref{FE3}) in 
apparently different forms. Recalling the definition of differential 
optical depth
\begin{equation}
\epsilon' = x_{\rm e} n_{0} \sigma_{\rm T} \frac{a}{a_{0}} = 
\frac{x_{\rm e}\, n_{0}\, \sigma_{\rm T}}{z + 1}.
\label{opdep}
\end{equation}
Defining with $\tau_{0}$ the conformal time at which the signal is received,
the optical depth will then be 
\begin{equation}
\epsilon(\tau,\tau_{0}) = \int_{\tau}^{\tau_{0}} x_{\rm e} n_{0} \sigma_{\rm T} 
\frac{a}{a_{0}} d\tau
\label{opdep2}
\end{equation}
while the visibility function \footnote{Recently the possible 
modifications induced by large-scale magnetic fields on the thermal history 
of the Universe during the ``dark age" beteween $z\sim 1000$ and $z\sim 10$
has been studied in \cite{ssethi}.} is 
\begin{equation}
{\cal K}(\tau) = \epsilon' e^{- \epsilon(\tau,\tau_{0})}.
\label{vis}
\end{equation}
Equation (\ref{vis}) gives the probability that a CMB photon was last 
scattered between $\tau$ and $\tau+ d\tau$ (usually in the literature 
${\cal K}(\tau)$ is denoted with $g(\tau)$).

Using Eq. (\ref{opdep2}),  Eq. (\ref{FE3}) can be written as 
\begin{equation}
\Omega_{\Phi}= \Phi' = \frac{d \Phi}{d \tau} = \frac{3 \,c^3\, \epsilon'}{ 16\pi^2 \, e} \vec{B}\cdot \hat{n},
\label{FE4}
\end{equation} 
where the explicit expression of the Thompson cross section in terms 
of the classical radius of the electron has been used, i.e. 
\begin{equation}
\sigma_{\rm T} = \frac{8\pi}{3} \,r_{0}^2,\qquad r_{0} = \frac{e^2}{m_{\rm e} c^2}.
\end{equation}
Equation (\ref{FE4})  can be also written in terms of the typical values 
of the magnetic field intensity and of the frequency of emission. Denoting 
with $\tilde{\nu}$ the physical frequency, i.e. $\tilde{\nu} = \nu/a$, it is clear that 
$\vec{\cal B}/\tilde{\nu}^2 = \vec{B}/\nu^2$ since the quantities in the numerator and in 
denominator redshift in the same way. Thus Eq. (\ref{FE4}) becomes:
\begin{equation}
\Omega_{\epsilon} =\frac{d \Phi}{d\epsilon} = 3.56\times 10^{-3} \,\, \biggl( \frac{ {\cal B}}{10^{-9}\,\,{\rm G}}\biggr) \,\, \biggl(\frac{ 100 \,\,\,
{\rm GHz}}{\tilde{\nu}}\biggr)^2 
\label{FE5}
\end{equation}
The typical quantities reported in Eq. (\ref{FE5}) refer to the {\em present} time. 
It is also conventional to use the same quantities but evaluated 
at the decoupling epoch. In this case the magnetic field will be of the order
of the ${\rm mG}$ while the physical frequency (at the decoupling time) 
will be of the order of $100\, {\rm THz}$.

\subsection{Faraday effect and radiative transfer equations}

The Faraday rotation rate derived in Eq. (\ref{FE4}) directly enters the 
radiative transfer equations. From Eq. (\ref{FE3}) it is possible 
to give an order of magnitude estimate of $\Phi$ itself.
Indeed, by integrating Eq. (\ref{FE3}) with respect to $t$, we can obtain that 
\begin{equation}
 \Phi= \frac{e^3}{ 2\, \pi\,c\, m_{\rm e}^2 \, \nu^2} 
\,\,\,\vec{B}\cdot \hat{n} \int x_{\rm e} \,\, n_{0} dt. 
\label{int1}
\end{equation}
Recalling that $ a(\tau) d \tau = d t$, the integral appearing in Eq. (\ref{int1}) 
can be easily estimated using Eq. (\ref{opdep2}); the result will be 
\begin{equation}
 \int x_{\rm e} \,\, n_{0} dt = \frac{\epsilon(\tau,\tau_{0})}{\sigma_{\rm T}}.
 \label{int2}
 \end{equation}
For orders of magnitude estimates, one can safely assume that the optical 
depth, i.e. $\epsilon(\tau,\tau_{0})$ is of order unity out to redshift of 
decoupling, when the polarization is generated \cite{FRT1,FRT2,FRT3}.  Therefore, after averaging 
over the polatizations of the magnetic field, the final result can be expressed 
in terms of $\sqrt{\langle \Phi^2 \rangle}$:
\begin{equation}
\sqrt{\langle \Phi^2 \rangle} = 1.6 \biggl(\frac{{\cal B}}{10^{-9} {\rm G}}\biggr) 
\biggl(\frac{30\,\, {\rm GHz}}{\tilde{\nu}}\biggr)^2 \,\,\, {\rm deg}. 
\end{equation}

To obtain more accurate estimates of the Faraday rotation rate, the relevant 
radiative transfer equations must be integrated \cite{FRT1} (see also 
\cite{cmbrev} for an introduction to the radiative transfer equations in CMB 
physics).  As it is well known, the polarization of the CMB is a higher 
order effect with respect to the temperature anisotropies and it depends
crucially upon the properties of the photon phase space distribution that 
should have a non-vanishing quadrupole moment. Indeed, to zeroth order 
in the tight-coupling expansion photons and baryons are synchronized so well 
that the CMB is not polarized. This means, technically, that the brightness 
perturbation associated with the Q Stokes parameter (i.e. $\Delta_{\rm Q}$) 
is zero to zeroth order in the tight coupling expansion. Since $\Delta_{\rm Q}$
measures the degree of linear polarization, the CMB, to this order, is not polarized.
However, to first order in the tight-coupling expansion a small amount of 
linear polarization is generated and it is proportional to the expansion 
parameter, i.e. $1/\epsilon'$, and to the (zeroth-order) dipole. 

This argument disregards the important fact that the Q and U Stokes 
parameters (and their associated brightness fluctuations 
$\Delta_{\rm Q}$ and $\Delta_{\rm U}$) are not 
invariant under rotations. It is possible to take into account 
this aspect by defining a generalized degree of polarization 
that is proportional to $(\Delta_{\rm U}^2 + \Delta_{\rm Q}^2)$.
 
 The presence of a magnetic field introduces then  further terms
 in the radiative transfer equations. These terms couple 
 directly the evolution of $\Delta_{\rm Q}$ to the evolution of 
  $\Delta_{\rm U}$ and the strength of the coupling is 
  given by the Faraday rotation rate $\omega_{\Phi}$ introduced in Eq. (\ref{FE5}).
  Therefore, the relevant radiative transfer equations can be written as
  \footnote{Here we are only discussing the Faraday effect induced on the 
  polarization in the case of scalar fluctuations of the geometry.}
\begin{eqnarray}
&& \Delta_{\rm Q}' +( i k\mu + \epsilon') \Delta_{\rm Q}  =   
\frac{\epsilon'}{2} [1- P_{2}(\mu)] S_{\rm Q} + 2 \Omega_{\Phi} \Delta_{\rm U},
\label{BRQ}\\
&& \Delta_{\rm U}' +( i k\mu  + \epsilon') \Delta_{\rm U}  =  - 2 \Omega_{\Phi} 
\Delta_{\rm Q},
\label{BRU}
\end{eqnarray}
where, 
\begin{equation}
S_{\rm Q}=  \Delta_{{\rm I}2} + \Delta_{{\rm Q}0} + \Delta_{{\rm Q}2}.
\label{source}
\end{equation}
Eqs. (\ref{BRQ}) and (\ref{BRU}) 
are derived by using the following expansions for the brightness 
perturbations:
\begin{eqnarray}
&& \Delta_{\rm Q}(\vec{k}, \hat{n}, \tau) = \sum_{\ell} (-i)^{\ell}\,\, (2 \ell + 1)\,\, \Delta_{{\rm Q}\,\ell} (\vec{k},\tau)\,\,P_{\ell}(\hat{k}\cdot\hat{n}),
\label{DQ1}\\
&& \Delta_{\rm U}(\vec{k}, \hat{n}, \tau) = \sum_{\ell} (-i)^{\ell}\,\, (2 \ell + 1)\,\, \Delta_{{\rm U}\,\ell} (\vec{k},\tau)\,\,P_{\ell}(\hat{k}\cdot\hat{n}),
\label{DQ2}
\end{eqnarray}
where $\vec{k}$  is the momentum of the Fourier expansion, $\hat{k}$ its direction; $ \hat{n}$ is the direction of the photon momentum and $\mu= \hat{k}\cdot \hat{n}$.
Consequently, in Eq. (\ref{source}), $\Delta_{{\rm I}2} $ denotes the quadrupole 
moment of the brightness fluctuation associated with the I Stokes 
parameter, while $\Delta_{{\rm Q}0}$ and  $\Delta_{{\rm Q}2}$ 
denote, respectively, the monopole and quadrupole moment of $\Delta_{\rm Q}$.
For completeness, and to anticipate further considerations 
on the cross power spectra, we also introduce the expansion for the brightness 
perturbation associated with the intensity of the radiation field:
\begin{equation}
 \Delta_{\rm I}(\vec{k}, \hat{n}, \tau) = \sum_{\ell} (-i)^{\ell}\,\, (2 \ell + 1)\,\, \Delta_{{\rm I}\,\ell} (\vec{k},\tau)\,\,P_{\ell}(\hat{k}\cdot\hat{n}).
\label{DI1def}
\end{equation}

An important remark concerns here the {\em symmetry} of the radiative transfer 
equations. In Eqs. (\ref{BRQ}) and (\ref{BRU}) $\mu$ denotes the 
projection of the Fourier mode along the direction of the 
photon momentum, i.e. $\mu = \hat{k} \cdot \hat{n}$. In Eqs. (\ref{BRQ}) 
and (\ref{BRU}) the magnetic field may introduce an extra preferred direction, so that 
the evolution equations will depend, ultimately, not only upon $\mu$ but also 
upon $\hat{B}\cdot\hat{n}$ (see Eq. (\ref{FE4})). Furthermore, the analysis 
may become even more cumbersome if the magnetic field is allowed not to be 
uniform (as assumed in the present and in the previous section). To make 
Eqs. (\ref{BRQ}) and (\ref{BRU}) more tractable the direction of the magnetic field 
intensity is taken to lie along the direction of the Fourier mode. 
Instead of working with $\Delta_{\rm Q}$ and $\Delta_{\rm U}$, it is 
useful to write down Eqs. (\ref{BRQ}) and (\ref{BRU}) directly 
in terms of the two associated complex conjugate combinations, i.e. 
\begin{equation}
{\cal M}_{\pm} = \Delta_{\rm Q} \pm i \Delta_{\rm U}. 
\label{CC}
\end{equation}
Using Eq. (\ref{CC}), Eqs. (\ref{BRQ}) and (\ref{BRU}) give, after some 
simple algebra, 
\begin{equation} 
{\cal M}_{\pm}' + ( i k \mu + \epsilon' \pm 2 i \Omega_{\Phi}) {\cal M}_{\pm} = 
\frac{3}{4} \epsilon' ( 1 - \mu^2) S_{\rm Q}.
\label{Mpm}
\end{equation}
Using the definitions given in Eqs. (\ref{opdep2}) and (\ref{vis}), Eq. (\ref{Mpm}) 
can be integrated easily and the result of the integration is 
\begin{equation}
{\cal M}_{\pm}(\vec{k}, \tau_{0}) = \frac{3}{4}( 1 -\mu^2) \int_{0}^{\tau_{0}} 
{\cal K}(\tau) e^{i k \mu (\tau-\tau_{0})} e^{\mp \,\epsilon(\tau,\tau_{0})
 \Omega_{\epsilon} (\hat{B}\cdot\hat{n}) } S_{Q}(\tau) \,\, d\tau,
 \label{Mpmsol}
 \end{equation}
 where $\Omega_{\epsilon}$ is given by Eq. (\ref{FE5}). This 
equation can be used for semi-analytical estimates of the mixing 
between $\Delta_{\rm Q}$ and $\Delta_{\rm U}$ to corroborate 
numerical estimates of the same effect. 

The analysis performed in terms of $\Delta_{\rm Q}$ and $\Delta_{\rm U}$ 
can also be expressed in terms of the E and B modes. 
In the latter language, the mixing between $\Delta_{\rm Q}$ and $\Delta_{\rm U}$ 
translates into a mixing between E and B modes \footnote{It should be clear that the $E$ abd B modes discussed here are not 
necessarily related to physical electric and magnetic fields.}.

Recalling, in fact,  that under clockwise rotations of $\varphi$
 ${\cal M}_{\pm}$ transform as
\begin{equation}
\tilde{{\cal M}}_{\pm} = e^{\mp 2i \varphi} {\cal M}_{\pm}
\end{equation} 
the combinations 
\begin{equation}
{\cal M}_{\pm}(\hat{n}) = \sum_{\ell m} a_{\pm 2,\ell m}
\,\,\,\,{_{\pm2}}{\cal Y}_{\ell m}(\hat{n})
\label{spin20}
\end{equation}
can be expanded in terms of the  spin-2 spherical
harmonics, i.e.   $_{\pm 2}{\cal Y}_{\ell}^{m}(\hat{n})$, with
\begin{equation}
a_{\pm 2,\ell m}=\int d\hat{n}\,\,\,\,\,\,{_{\pm2}} {\cal Y}_{\ell m}^{\ast}(\hat{n})\,\,(\Delta_{\rm Q} \pm i \Delta_{\rm U})(\hat{n}).
\label{spin21}
\end{equation}
The ``electric" and ``magnetic" components of polarization are 
eigenstates of parity and may be defined as 
\begin{equation}
a_{\ell m}^{\rm E}=-\frac{1}{2}(a_{2, \ell m}+a_{-2,\ell m}),\qquad
a_{\ell m}^{\rm B}= \frac{i}{2}(a_{2,\ell m}-a_{-2,\ell m}).
\label{defaEB}
\end{equation}
These newly defined variables are expanded in terms of ordinary spherical
harmonics, $Y_{\ell m}(\hat{n})$,
\begin{equation}
{\rm E}(\hat{n})= \sum_{\ell m} a_{\ell m}^{\rm E} Y_{\ell m}(\hat{n}), \,\,\,\,\,\,\,\,\,\,\,\,\,
{\rm B}(\hat{n})= \sum_{\ell m} a_{\ell m}^{\rm B} Y_{\ell m}(\hat{n}).
\label{EB}
\end{equation}
In connection with the spin-2 spherical harmonics  appearing in Eqs. (\ref{spin20})
and (\ref{spin21}) it is relevant to mention here that a generic spin-s spherical 
harmonic is a known concept in the quantum mechanical theory 
of angular momentum (see, for instance, \cite{sakurai}). A typical 
quantum mechanical problem is to look for the representations
of the operator specifying three-dimensional rotations, i.e. $\hat{R}$; 
this problem  is usually approached within the so-called Wigner matrix elements, i.e. 
${\cal D}_{m \, m'}^{(j)}(R) = \langle j, \, m'| \hat{R} | j,\,m>$ where 
$j$ denotes the eigenvalue of $J^2$ and $m$ denotes the eigenvalue of $J_{z}$. 
Now, if we replace $m' \to - {\rm s}$, $j\to \ell$, we have the definition 
of spin-s spherical harmonics in terms of the 
${\cal D}_{ -{\rm s},\, m}^{(\ell)}(\alpha,\beta, 0)$, i.e.  
\begin{equation}
_{\rm s}{\cal Y}_{\ell\, m}(\alpha,\beta) = \sqrt{\frac{2\ell +1}{4\pi}} {\cal D}_{-{\rm s},\, m}^{(\ell)}(\alpha,\beta, 0),
\label{spin23}
\end{equation}
where $\alpha$, $\beta$ and $\gamma$ (set to zero in the above definition) 
are the Euler angles defined as in \cite{sakurai}. If $s=0$, ${\cal D}_{0,\, m}^{(\ell)}(\alpha,\beta, 0) = \sqrt{(2\ell +1)/4\pi} Y_{\ell \, m}(\alpha,\beta)$ where 
$Y_{\ell \, m}(\alpha,\beta)$  are the ordinary spherical harmonics.

To define properly the cross power spectra we also recall the expansion 
of the intensity fluctuations of the radiation field (defined in Eq. (\ref{DI1def})),
in terms of spherical harmonics:
\begin{equation}
\Delta_{\rm I}(\hat{n}) = \sum_{\ell m} a_{\ell m}^{\rm T} Y_{\ell m}(\hat{n}), 
\label{T}
\end{equation}
where we wrote explicitly $a_{\ell m}^{\rm T}$ since the fluctuations 
in the intensity of the radiation field, i.e.  $\Delta_{\rm I}$ are nothing 
but the fluctuations in the CMB temperature. 

Under parity inversion,  the components appearing in Eqs. (\ref{EB}) and 
(\ref{T}) transform as 
\begin{equation}
 a_{\ell m}^{\rm E} \to (-1)^{\ell }\,\,a_{\ell m}^{\rm E},\qquad
 a_{\ell m}^{\rm B} \to (-1)^{\ell +1 }\,\,a_{\ell m}^{\rm B},\qquad
 a_{\ell m}^{\rm T} \to (-1)^{\ell }\,\,a_{\ell m}^{\rm T}.
\label{transform}
\end{equation}
Therefore, the E-modes have the same parity of the T-modes 
which have, in  turn, the same parity of spherical harmonics, i.e. 
$(-1)^{\ell}$. On the contrary, the B-modes have $(-1)^{\ell + 1}$ parity.

The existence of linear polarization allows for 6 different cross power 
spectra to be determined, in principle, from data that measure 
the full temperature and polarization anisotropy information.
The cross power spectra can be defined in terms of the spectral functions 
$C_{\ell}^{X,Y}$ where $X$ and $Y$ stand for E, B or T depending 
on the cross-correlation one is interested in:
\begin{equation}
C_{\ell}^{X,Y} = \frac{1}{2 \pi^2} \int k^2 \, dk \sum_{m = -\ell}^{\ell} \frac{( a_{\ell m}^{X})^{\ast} a_{\ell m}^{Y}}{( 2 \ell + 1)}.
\label{defcross}
\end{equation}
Therefore, if we are interested in the TE correlations we just have to set 
$X= {\rm T}$ and $Y ={\rm E}$ and use the relevant expansions given above. 
In the following, we will denote the correlations as TT, EE, BB, TB and so on.
This notation refers to the definition given in Eq. (\ref{defcross}).

Let us now see how the definition (\ref{defcross}) works. Suppose 
we are interested in the TT correlations, i.e. the usual and well 
known temperature correlations. From Eq. (\ref{defcross}) we will have 
\begin{equation}
C_{\ell}^{{\rm TT}} = \frac{1}{2 \pi^2} \int k^2 \, dk \sum_{m = -\ell}^{\ell} 
\frac{[a_{\ell m}^{{\rm T}}(k)]^{\ast} a_{\ell m}^{{\rm T}}(k)}{( 2 \ell + 1)}.
\label{defcross2}
\end{equation}
Now, from Eq. (\ref{T}), using the orthogonality of spherical harmonics, we have 
that 
\begin{equation}
a_{\ell m}^{\rm T}(k) = \int d \hat{n} Y_{\ell m}(\hat{n}) \Delta_{\rm I}(\vec{k}, \hat{n}).
\label{inverse}
\end{equation}
Inserting Eq. (\ref{inverse}) into Eq. (\ref{defcross2}) and recalling 
the expansion of $\Delta_{\rm I}(\vec{k}, \hat{n})$ in terms 
of Legendre polynomials we get 
\begin{equation}
C_{\ell}^{\rm TT} = \frac{2}{\pi} \int dk \, k^2 |\Delta_{{\rm I}\,\ell}|^2.
\label{Cell}
\end{equation}
To get to Eq. (\ref{Cell}) the following two identities 
have been used, i.e. 
\begin{eqnarray}
&& \int d\hat{n} P_{\ell'}(\hat{k}\cdot\hat{n}) Y_{\ell m}^{\ast}(\hat{n}) = 
\frac{4\pi}{(2 \ell + 1)} \delta_{\ell\ell'},
\nonumber\\
&& \sum_{m  = -\ell}^{\ell} Y_{\ell m}^{\ast}(\hat{k}) Y_{\ell m}(\hat{n}) =
\frac{2\ell + 1}{4\pi} P_{\ell}(\hat{k}\cdot\hat{n}).
\label{addition}
\end{eqnarray}
The second identity in Eq. (\ref{addition}) has been already exploited 
in Eq. (\ref{ADD}).

In similar ways, different expressions for the other correlations 
may be obtained. Notice that Eq. (\ref{Cell}) is a consequence 
of the specific conventions adopted in Eq. (\ref{DI1def}).
In particular note that the a factor $(2 \ell +1)$ is included 
in the expansion. It must be clearly said that this is matter 
of conventions. For  instance, in 
Refs. \cite{cmbrev} and \cite{MB} the factor $(2\ell +1)$ as well as the factor 
$(-i)^{\ell}$ are included in the expansion \footnote{Notice that the authors 
of Ref. \cite{FRT3} do not include the factor $(-i)^{\ell}$ but do include 
the factor $(2\ell +1)$ in the expansion of the various brightness 
perturbations.}. On the contrary, in Ref. \cite{HS1}, the authors did not include 
the factor $(2\ell +1)$ in the expansions for the brightness perturbations 
that we defined in Eqs. (\ref{DQ1}), (\ref{DQ2}) and (\ref{DI1def}).
Consequently, the cross correlations will inherit extra-terms that are 
simply a consequence of the different conventions adopted. So, for instance, 
in Eq. (\ref{Cell}), and in the conventions of \cite{HS1}, a factor $(2\ell +1)^2$ 
typically appears in the denominator.

Recalling the connection between the Wigner matrix elements and the spin-s 
spherical harmonics, i.e. Eq. (\ref{spin23}), it is possible to show that, under 
complex conjugation, 
\begin{equation}
a_{\pm 2,\ell m}^{\ast} = (-1)^{m} a_{\mp 2,\ell m},\qquad \big(a_{\ell, m}^{{\rm T}, {\rm E}, {\rm B}} \bigr)^{\ast} = (-1)^{m}a_{\ell, - m}^{{\rm T}, {\rm E}, {\rm B}},
\label{CC1} 
\end{equation}
 where the second equality follows from the first one by using Eq. (\ref{defaEB}).
 It is then possible to show that while the TB and EB correlators are parity-odd, 
 all the other correlators (i.e. TT, BB, EE, TE) are parity even. 
\subsection{Limits on uniform magnetic fields from Faraday effect}
In the absence of magnetic fields, the scalar fluctuations of the 
geometry only generate E-modes. The presence of a magnetic field 
induces, however, a mixing between E-modes and B-modes. 
Consequently, if we have an initial E-mode generated by scalar 
metric perturbations, the presence of a magnetic field induces a non-vanishing 
B-mode.  In the case of scalar fluctuations, a non-vanishing TE
power spectrum arises directly as a result of the CMB polarization 
as it can be verified by studying the first-order tight coupling expansion. 
Recently, the WMAP collaboration \cite{wmap1,wmap2} measured 
the TE correlations. 
\begin{figure}
\centerline{\epsfxsize = 11cm  \epsffile{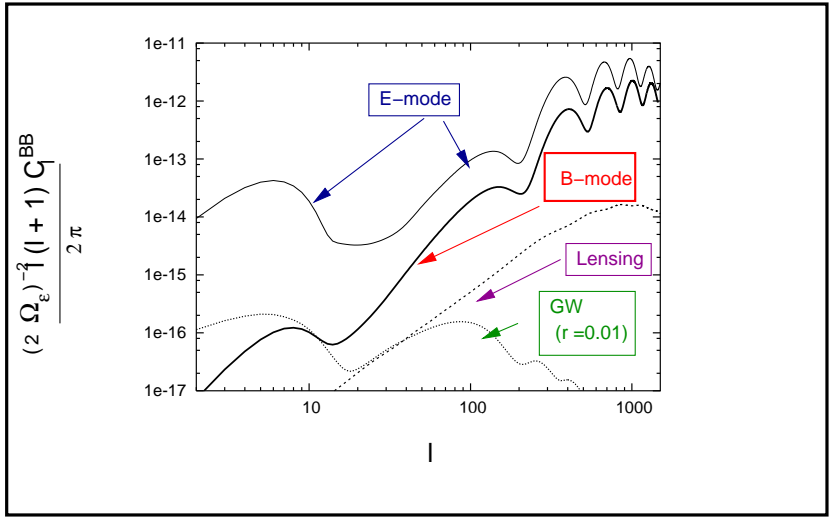}}
\vskip 3mm
\caption[a]{Angular autocorrelation of B-modes of CMB polarization generated by a uniform magnetic field in a cosmic concordance model 
(adapted from Ref. \cite{scocco}). Notice that the vertical axis 
of has been rescaled by $(2 \Omega_{\epsilon})^{-2}$ (see Eq. (\ref{FE5})).
With the dashed line marked with ``lensing" the possible B-mode arising from 
weak lensing is reported. The B-mode induced by GW is also reported (dotted line)
for a typical value of the tensor to scalar ratio $r = 0.01$.}
\label{FIG4} 
\end{figure}
As discussed in \cite{scocco},  the induced 
TB angular power spectrum is, roughly $\Omega_{\epsilon}$ times 
smaller than the initial $TE$ correlations.  From the WMAP measurement, 
the measured value of the $TE$ correlation and the absence 
of the TB correlations allows to set a limit on $\Omega_{\epsilon}$. At 
a given frequency, the limit on $\Omega_{\epsilon}$ can be 
translated into a bound on the magnetic field intensity, i.e. 
\begin{equation}
{\cal B}_0 < 10^{-8}\,\,{\rm G}
\end{equation}
at a typical (present) frequency of $30$ GHz \cite{scocco}.
The values of the TE and EE correlations induced by the presence of a 
magnetic field are $\Omega_{\epsilon}^2$ times smaller than their values 
in the absence of the magnetic fields. 
In Fig. \ref{FIG4} the autocorretaion of the $B$-modes of CMB polarization 
is reported in the case of a uniform magnetic field (recall that $\Omega_{\epsilon}$ is defined in Eq. (\ref{FE5})).  Notice that when 
$ 2 \Omega_{\epsilon} \simeq 1$ the $TE$ (as well as the the other correlations) are strongly damped and the differential Faraday rotation 
has a net depolarizing effect \cite{FRT3}. 

We conclude this section by remarking that Faraday rotation 
is not the only way to generate $B$-modes. In fact:
\begin{itemize}
\item{} B-modes can be generated {\em directly} by fully inhomogeneous 
magnetic fields (see following sections);
\item{} B-modes can be generated by gravitational waves (in Fig. \ref{FIG4})
the GW contribution is reported for a value of the tensor-to scalar ratio 
compatible with the present WMAP data);
\item{} B-modes can be generated by weak lensing.
\end{itemize}
To this list one should also add the possibility that some other form 
of birefringence is allowed 
(as elaborated in the derivation of Eq. (\ref{GENFAR})). In connection 
with the GW contribution, it should be noted that $r$ (denoting the 
ration between 
the tensor and scalar (primordial) power spectra)  is limited from above, by 
current experimental data, i.e. $ r < 0.53$. In Fig. \ref{FIG4} the contribution 
of the B-mode induced by GW is illustrated in the case $r =0.01$.

The peculiar feature of the Faraday rotation signal 
induced by a uniform magnetic field is the frequency dependence.
In fact, by going at higher frequencies, the signal gets reduced as $\nu^{-2}$.
On the contrary the B-modes induced by gravitational waves does not depend upon the frequency. 
The birefringence induced by a dynamical quintessence (pseudo-scalar) field 
is also independent on the frequency. The need of multifrequency determination of Faraday rotation is then clear and this point has been 
also correctly emphasized in \cite{dolgov1,dolgov2}. In \cite{dolgov1}
a scheme for the treatment of Faraday rotation
in the case of fully inhomogeneous magnetic fields has been 
studied and along a similar line another analysis  has also been 
presented in \cite{tk3}.

In the following years on the CMB side 
the Planck mission \cite{planck} will give us, hopefully, important clues 
concerning CMB polarization. For the purposes of the present 
subject it is appropriate to recall the Planck will operate in 9 frequency channels ranging from $30$ to $857$ GHz.  Multifrequency 
Faraday rotation measurements may become possible. 
Unfortunately, owing to the fact that the nominal Planck sensitivity to 
polarization is larger at higher frequencies, these measurements may also be 
difficult. In fact, Faraday rotation signals are suppressed by the square 
of the frequency and are therefore smaller at high frequencies.
Other instruments dedicated to the analysis of CMB polarization 
may also give important clues. These experiments include 
polarization-sensitive balloon borne experiments (like 
the polarization-sensitive version of Boomerang \cite{boomerang,masi} and its 
possible improvements)
as well as ground based arrays 
like the promising QUIET (Q/UImagingExperimenT) \cite{quiet}. 
On the radio-astronomical side, the very ambitious SKA (Square Kilometer 
Array) project will be able to give us precious full-sky surveys 
of Faraday rotation measurements \cite{SKA} (see also \cite{govoni,feretti}).
The SKA experiment should cover a frequency range between $0.1$ and $20$ 
GHz.

\renewcommand{\theequation}{4.\arabic{equation}}
\section{One-fluid plasma description}
\setcounter{equation}{0}

Since a fully inhomogeneous magnetic field acts as a source term
in the evolution equations of the gravitational perturbations, the CMB 
observables may be affected. 
To investigate this situation, the relevant 
system of equations describing the large-scale magnetic fields is 
slightly different from the one introduced in Eqs. (\ref{div1}), (\ref{bianchi1}) 
and (\ref{current}) (where, however, the displacement 
current may be neglected). In fact, we are here interested in the dynamics 
taking place for typical frequencies much smaller than  $\omega_{\rm p,\,\,e}$ and 
$\omega_{\rm p,\,\,i}$. In this situation, from the two-fluid equations of 
Eqs.  (\ref{div1}), (\ref{bianchi1}) 
and (\ref{current}) it is possible to derive various one-fluid plasma 
descriptions by defining a set of appropriate one-fluid variables 
such like the total current  \cite{biskamp,krall}
\begin{equation}
\vec{J} = e ( n_{i} \vec{v}_{\rm i} - n_{\rm e} \vec{v}_{\rm e}),
\end{equation}
the center-of-mass velocity 
\begin{equation}
\vec{v} = \frac{ m_{\rm i} \vec{v}_{\rm i} + m_{\rm e} \vec{v}_{\rm e}}{m_{\rm e} + m_{\rm i}},
\end{equation}
and so on. Among the one-fluid descriptions, a particularly simple 
possibility is provided by magnetohydrodynamics (MHD) that can be 
characterized by the following set of conditions:
\begin{equation}
\vec{\nabla}\cdot \vec{E} =0,\qquad \vec{\nabla}\cdot \vec{B}=0,\qquad
\vec{\nabla}\cdot \vec{J} =0.
\label{solenoidal}
\end{equation}
As mentioned above, the derivative of the electric field is neglected in MHD analog of Eq. (\ref{current}). Moreover, since the current is solenoidal, as the magnetic field 
intensity, it follows, from Maxwell's equations, that 
\begin{equation}
\vec{J} = \frac{1}{4\pi} \vec{\nabla}\times \vec{B}.
\label{currmhd}
\end{equation}
The induced (Ohmic) electric field is therefore given, in the absence of Hall
and thermoelectric terms, by 
\begin{equation}
\vec{E} + \vec{v}\times \vec{B} = \frac{1}{4\pi\sigma} \vec{\nabla}\times \vec{B},
\label{ohmic}
\end{equation}
where $\sigma$ is the conductivity and where we used 
units where the speed of light is equal to 1. In a relativistic plasma 
$\sigma \simeq T/\alpha_{\rm em}$ while in a non-relativistic 
plasma $\sigma \simeq (T/\alpha_{\rm em}) (T/m_{\rm e})^{1/2}$. 
Since the conductivity increases as we go back in time, in the 
early Universe the condition $\vec{E} + \vec{v}\times\vec{B}/c\simeq 0$ 
is approximately verified. 
Under these conditions, two (approximate) conservations laws 
may be derived, namely the magnetic flux conservation 
\begin{equation}
\frac{d}{d \tau} \int_{\Sigma} \vec{B} \cdot d\vec{\Sigma}=-
\frac{1}{4\pi \sigma} \int_{\Sigma} \vec{\nabla} \times\vec{\nabla}
\times\vec{B}\cdot d\vec{\Sigma},
\label{flux}
\end{equation}
and the magnetic helicity conservation
\begin{equation}
\frac{d}{d\tau} \biggl(\int_{V} d^3 x \vec{A}~\cdot \vec{B}\biggr) = - \frac{1}{4\pi \sigma} \int_{V} d^3 x
{}~\vec{B}\cdot\vec{\nabla} \times\vec{B}.
\label{hel}
\end{equation}
In Eq. (\ref{flux})  $\Sigma$ is an arbitrary closed surface that moves with the
plasma. According to Eq. (\ref{flux}),  in MHD the magnetic field 
has to be always solenoidal (i.e. $\vec{\nabla} \cdot \vec{B} =0$).
Thus, the magnetic flux conservation implies that, in the 
superconducting limit (i.e. $\sigma \to \infty$) the 
magnetic flux lines, closed because of the transverse nature of the field, evolve always 
glued together with the plasma element. In this 
approximation, as far as the magnetic field 
evolution is concerned, the plasma is a collection 
of (closed) flux tubes. The theorem of flux conservation 
states then  that the energetical properties of large-scale
magnetic fields are conserved throughout the plasma evolution. 

While 
the flux conservation concerns the 
energetic properties of the magnetic flux lines, the 
magnetic helicity, i.e. Eq. (\ref{hel}), concerns chiefly the 
 topological  properties of the 
magnetic flux lines. In the simplest situation,
 the magnetic flux lines will be closed loops 
evolving independently in the plasma and the helicity 
will vanish. There could be, however, 
more complicated topological situations \cite{mgknot}
where a single magnetic loop is twisted (like some 
kind of M\"obius stripe) or the case where 
the magnetic loops are connected like the rings of a chain:
now the non-vanishing magnetic helicity 
 measures, essentially, the number of links and twists 
in the magnetic flux lines \cite{biskamp}. Furthermore, in the 
superconducting limit, the helicity will not change 
throughout the time evolution. 
The conservation of the magnetic flux and of the magnetic 
helicity is a consequence of the fact that, in ideal 
MHD, the Ohmic electric field is always orthogonal 
both to the bulk velocity field and to the magnetic 
field. In the resistive MHD approximation this conclusion may  not apply
\footnote{The presence (or absence) of magnetic helicity plays an important 
r\^ole for the possibility of inverse cascades in the pre-recombination plasma. In particular, one of the 
evolutions of the correlation scale presented in Fig. \ref{FIG1} refers to the case 
where the magnetic field is helical \cite{son}.}

The quantity at the right-hand-side of Eq. (\ref{hel}), i.e. $\vec{B}\cdot\vec{\nabla}\times \vec{B}$ is called magnetic gyrotropy and it is a gauge-invariant measure 
of the number of contact points in the magnetic flux lines.  The magnetic 
helicity (or the magnetic gyrotropy) and the Lorentz force can be 
used to give a general classification of the possible inhomogeneous
magnetic field configurations that are relevant in a MHD context. 
Since the current density is solenoidal, we have that, in MHD, 
\begin{equation}
\vec{F}_{\rm B} = \vec{J} \times \vec{B} \simeq \frac{1}{4\pi} (\vec{\nabla}\times 
\vec{B})\times \vec{B}.
\label{LF}
\end{equation}
Looking at the expressions of the magnetic gyrotropy and of the 
Lorentz force, there are, in principle, two extreme situations 
that may arise:
\begin{itemize}
\item{} in the first case $\vec{\nabla}\times\vec{B}\times \vec{B}=0$ 
and the magnetic gyrotropy, i.e. $\vec{B}\cdot \vec{\nabla}\times \vec{B}$, is maximal;
\item{} in the second case the Lorentz force is maximal and the magnetic 
gyrotropy vanishes.
\end{itemize}

In the case when the Lorentz force term vanishes, the configuration is said to be 
force-free. In this limit the MHD equations (coupled with the Navier-Stokes equation)
become particularly simple and appropriate for the deduction of 
useful scaling laws \cite{biskamp,olesen1,olesen2,olesen3}. 
Fully inhomogeneous magnetic field configurations 
may then gravitate and affect the evolution equations of the 
tensor, vector and scalar modes of the geometry. Therefore, in the following section,
the evolution equations of metric fluctuations will be investigated in the 
presence of fully inhomogeneous magnetic fields.
 MHD equations in curved space-times are also relevant for 
 a number of different problems arising in the physics of large-scale 
 magnetic field. Concerning the various aspects of this subject 
 see, for instance,  Refs. \cite{enqvist1,review2,widrow} and references therein. 

As a last comment, we would like to stress that the magnetic helicity 
as well as the magnetic gyrotropy are parity-odd quantities. Therefore, recalling the 
considerations related to Eq. (\ref{CC1}), if 
sources of gyrotropy and helicity are present after equality the 
BE and TB correlators may not be zero. This point has been 
discussed in \cite{pogosian} (see, however, also \cite{dolgov1}).
\subsection{Fully inhomogeneous magnetic correlators}

Nearly all 
mechanisms able to generate large scale magnetic fields imply the 
existence of a stochastic background of magnetic disturbances that 
could be written, in Fourier space, as \footnote{For the Fourier 
transforms we use the following conventions: $B_{i} (\vec{x}) = 
(2\pi)^{-3} \,\,\int d^{3}k  e^{- i \vec{k}\cdot\vec{x}} B_{i}(\vec{k})$
and, conversely, $B_{i} (\vec{k}) = 
\int d^{3}x  e^{ i \vec{k}\cdot\vec{x}} B_{i}(\vec{x})$.}
\begin{equation}
\langle B_{i}(\vec{k},\tau) B_{j}^{\ast}(\vec{p},\tau) \rangle = (2\pi)^3 P_{ij}(k)  \delta^{(3)}(\vec{k} - \vec{p}),
\label{stoch1}
\end{equation}
where 
\begin{equation}
P_{ij}(k) = {\cal Q}(k) \biggl(\delta_{ij} - \frac{k_{i}\, k_{j}}{k^2}\biggr),\qquad 
{\cal Q}(k) = {\cal Q}_{0} k^{m}.
\label{definitions1}
\end{equation}
From Eq. (\ref{definitions1})  the magnetic field configuration 
of Eq. (\ref{stoch1}) depends on the amplitude of the field ${\cal Q}_{0}$ and on 
the spectral index $m$. 

It is easy to check that, in the case of the configuration (\ref{stoch1}) 
the magnetic gyrotropy is vanishing, i.e. 
\begin{equation}
\langle \vec{B} \cdot \vec{\nabla} \times \vec{B} \rangle =0, 
\qquad \langle (\vec{B} \cdot \vec{\nabla} \times \vec{B})^2 \rangle =0,
\end{equation}
where the second expression denotes, for short, the two-point 
function of the magnetic gyrotropy. On the 
contrary it should also be clear that, from Eq. (\ref{stoch1}), that 
$\langle(\vec{\nabla}\cdot[\vec{\nabla}\times \vec{B})\times \vec{B}] \rangle \neq 0$.

There are situations where magnetic fields are produced in a 
state with non-vanishing gyrotropy (or helicity) \cite{q4,q4a,pogosian,tk2}.
In this case, the two point function can be written in the same 
form given in Eq. (\ref{stoch1}) 
\begin{equation}
\langle B_{i}(\vec{k},\tau) B_{j}^{\ast}(\vec{p},\tau) \rangle = (2\pi)^3 \tilde{P}_{ij}(k)  \delta^{(3)}(\vec{k} - \vec{p}),
\label{stoch2}
\end{equation}
but where now 
\begin{equation}
\tilde{P}_{ij}(k) = {\cal Q}(k) \biggl(\delta_{ij} - \frac{k_{i}\, k_{j}}{k^2}\biggr) + i 
\tilde{{\cal Q}}(k) \epsilon_{i j \ell} \,\frac{k^{\ell}}{k},\qquad 
\tilde{{\cal Q}}(k) = \tilde{{\cal Q}}_{0} k^{\tilde{m}}.
\label{definitions2}
\end{equation}
From Eq. (\ref{definitions2}) we can appreciate that, on top of the parity-invariant 
contribution (already defined in Eqs. (\ref{stoch1}) and (\ref{definitions1})), there 
is a second term proportional to the Levi-Civita $\epsilon_{ij\ell}$.
In Fourier space, the introduction of gyrotropic configurations 
implies also the presence of a second function of the momentum 
$\tilde{{\cal Q}}(k)$.

As it will become more plausible from the considerations reported 
in Section 5, the two-point function 
determines, eventually, the higher-order correlation functions. 
Since the contribution of fully inhomogeneous magnetic fields
 to the energy-momentum tensor is quadratic, 
 it is rather frequent, in various calculations, to evaluate four-point correlation 
functions.  Fourier space is not always the best framework for 
the evaluation of higher-order correlation functions. Indeed, a swifter derivation 
of various results can be achieved, in some cases, by staying in real space (and by 
eventually Fourier transforming the final result). Let us 
therefore elaborate on the real space correlators associated with 
the configurations discussed in Eqs. (\ref{stoch1}) and 
(\ref{stoch2}).  

The real space  two-point function can be written, for instance, as
\begin{equation}
{\cal G}_{ij}(r) = \langle B_{i}(\vec{x} + \vec{r}) B_{j} (\vec{x})\rangle 
= {\cal F}_{1}(r) \delta_{ij} + {\cal F}_{2}(r) r_{i} r_{j} + 
{\cal F}_{3}(r) \epsilon_{ijk}r^{k}. 
\label{rsc}
\end{equation}
If ${\cal F}_{3}(r) =0$ then we are in the situation described by Eq. (\ref{stoch1}).
If ${\cal F}_{3}(r) \neq 0$ the framework is the one defined by Eq. (\ref{stoch2}).

Consider, for instance, the case ${\cal F}_{3}(r) =0$.
Then, using the parametrization 
(\ref{rsc}), Eqs. (\ref{stoch1}) and (\ref{definitions1}) lead to 
\begin{equation}
{\cal G}_{ij}(r)={\cal G}_{\perp}(r) \delta_{ij}
+  [ {\cal G}_{\parallel}(r) - {\cal G}_{\perp}(r)] \frac{r_{i} r_{j}}{r^2},
\label{rsc2}
\end{equation}
where 
\begin{eqnarray}
&&{\cal G}_{\perp}(r) = \frac{4}{(2\pi)^2} \int k^2 d k \,\,{\cal Q}(k) \biggl[ j_{0}(kr) - 
\frac{j_{1}(kr)}{kr}\biggr],
\nonumber\\
&& {\cal G}_{\parallel}(r) =  \frac{8}{(2\pi)^2}
 \int k^2 d k \,\,{\cal Q}(k) \frac{j_{1}(kr)}{kr},
 \end{eqnarray}
 where $j_{0}(k r)$ and $j_{1}(k r)$ are the usual spherical Bessel functions 
 \cite{abr,grad}:
 \begin{equation}
 j_{0}( k r) = \frac{\sin{k r}}{k r},\qquad j_{1}(k r) = \frac{\sin{k r}}{(k r)^2} - 
 \frac{\cos{k r}}{k r}.
 \end{equation}
 
Another useful parametrization of the real space correlators is the 
one where the functions $F_{1}(r)$ and $F_{2}(r)$ are written as derivatives 
of a function $f(r^2)$ whose asymptotic behaviour can be 
determined from the trace of the two-point function \cite{q4}. 
More precisely, in the case $F_{3}(r)=0$, the functions 
appearing in Eq. (\ref{rsc}) can be writtten as
\begin{equation}
{\cal F}_{1}(r) = \frac{\partial}{\partial r^2} [ r^2 f(r^2)],\qquad 
{\cal F}_{2}(r) = - \frac{\partial}{\partial r^2} [ f(r^2)].
\label{definitions3}
\end{equation}
With the definitions (\ref{definitions3}), Eq. (\ref{rsc}) is clearly 
transverse (i. e. $\partial^{i} {\cal G}_{ij} =0$). Moreover, defining 
as ${\cal G}(r) = {\rm Tr}[{\cal G}_{ij}]$, we also have 
\begin{equation}
r {\cal G}(r) = 2 \frac{\partial}{\partial r^2} [r^3 f(r^2)].
\end{equation}

It is often  useful, in practical estimates, to regularize the two-point function 
by using an appropriate ``windowing" \cite{maxmis}. 
Two popular windows  
are, respectively, the Gaussian and the top-hat functions , i.e.
\begin{equation}
{\cal W}_{\rm g}(k,L) = e^{- \frac{k^2 L^2}{2}},\qquad {\cal W}_{\rm th}(k,L) =
\frac{3}{k L}\,j_{1}( k L).
\end{equation}

For instance, the regularized trace of ${\cal G}_{ij}(r)$ with Gaussian 
filter can be obtained from the previous expressions by shifting 
${\cal Q}(k) \to {\cal Q}(k) W_{\rm g} (k, L)$. The result is 
\begin{equation}
{\cal G}(r) = \frac{ 2 {\cal Q}_{0}}{( 2\pi)^2} \frac{1}{L^{3 + m}} F\biggl( \frac{m + 3}{2}, \frac{3}{2}, - \frac{r^2}{4 L^2} \biggr),
\label{trace}
\end{equation}
where $F(a, b, x) \equiv_{1} F_{1}(a,b, x) $ is the confluent hypergeometric function \cite{abr,grad}. Notice that the integral appearing 
in the trace converges for $m > -3$.  The amplitude of the magnetic power 
spectrum ${\cal Q}_{0}$ can be traded for ${\cal G}(0) \equiv B^2_{L}$ (as 
suggested, for instance, in \cite{tk1}). 
From Eq. (\ref{trace}) we have that ${\cal Q}_{0}$ becomes 
\begin{equation}
{\cal Q}_{0} = \frac{(2 \pi)^{ m + 5}}{2} \frac{k_{L}^{-( 3 + m)}}{\Gamma\biggl(
\frac{m + 3}{2}\biggr)} B_{L}^2,
\label{Q0}
\end{equation}
 where $k_{L} = 2\pi/L$.
 
 In the real space approach the higher order correlators are reduced to the 
 calculation of derivatives of special functions (depending upon the 
 regularization scheme). For instance, we can compute easily, in general terms 
 the correlation function for the energy density fluctuations. 
 Defining 
 \begin{equation}
 \delta \rho_{B}(\vec{x}) = \frac{1}{8\pi} B_{i}(\vec{x}) B^{i}(\vec{x}),
 \end{equation}
 it is sometimes important to evaluate 
 \begin{equation}
 {\cal E}(r) = \langle \delta\rho_{B}(\vec{x}) \delta\rho_{B}(\vec{y}) \rangle
 - \langle \delta\rho_{\rm B}(\vec{x}) \rangle \langle \delta\rho_{B}(\vec{y}) \rangle.
 \end{equation}
 By using the decomposition (\ref{rsc}) we will have (taking for simplicity 
 the case $F_{3}(r)=0$), 
 \begin{eqnarray}
 {\cal E}(r) &=& \frac{1}{32 \pi^2}[ 3 F_{1}^2(r) + 2 r^2 F_{1}(r) F_{2}(r) + r^4 
 F_{2}^2(r) 
 \nonumber\\
 &=& \frac{1}{32 \pi^2} \biggl[ 3 f^2(r^2) + 4r^2 f(r^2) \frac{\partial f}{\partial r^2} 
 + 2 r^4 \biggl(\frac{\partial f}{\partial r^2}  \biggr)^2 \biggr],
 \end{eqnarray}
 where the second equality follows from Eq. (\ref{definitions3}).
 
With these (or similar techiques) higher order correlation functions 
for magnetic inhomogeneities may be computed for the interesting 
observables like the Lorentz force \cite{tk1}, the magnetic gyrotropy \cite{q4} and so on. 

It is appropriate to comment here 
on the parametrizations usually employed in the analysis of the various mechanisms for the magnetic field production in the early Universe. 
This topic (together with other related subjects) has been reviewed in  
Ref. \cite{review2}, and, consequently that discussion will not 
be repeated here. For sake of simplicity consider 
the mechanisms  relying on super-adiabatic amplification 
of vacuum fluctuations of the gauge fields (see also Fig. \ref{FIG2}).
In this context what one is led to compute is the amplification of the 
magnetic field from some initial (vacuum) state. The precise 
amount of amplification (as well as the spectral properties 
of the final configuration) is determined by the features of the 
pump field that changes from model to model.
Therefore, initially (i.e. for $\tau \to -\infty$) 
the quantum mechanical operator 
related to the magnetic field intensity can be decomposed as 
\begin{equation}
\hat{B}^{\rm in}_{i}(\vec{x},\tau) = \frac{i \epsilon_{m n i}}{(2\pi)^3} 
\sum_{\alpha} e_{n}^{\alpha} \int d^{3} k\,\, k_{m} [ \hat{a}_{\vec{k},\alpha}
f_{k}(\tau) e^{- i \vec{k} \cdot \vec{x}} +
 \hat{a}^{\dagger}_{\vec{k},\alpha} f_{k}^{\ast}(\tau)
e^{ i \vec{k} \cdot\vec{x}}],
\label{mdec1}
\end{equation}
where, within the present conventions, $[\hat{a}_{\vec{k},\alpha}, 
\hat{a}^{\dagger}_{\vec{p,\beta}}] =  (2\pi)^3 \delta_{\alpha\beta} \delta^{(3)}(\vec{k} - \vec{p})$.
For $\tau\to +\infty$ 
\begin{equation}
\hat{B}^{\rm out}_{\ell}(\vec{x},\tau) = \frac{i \epsilon_{a b \ell}}{(2\pi)^3} 
\sum_{\beta} e_{b}^{\beta}\int d^{3}k\, \, k_{a} [ \hat{b}_{\vec{k},\beta}
g_{k}(\tau) e^{-i \vec{k} \cdot \vec{x}} +
 \hat{b}^{\dagger}_{\vec{k},\beta} g_{k}^{\ast}(\tau)
e^{ i \vec{k} \cdot\vec{x}}].
\label{mdec2}
\end{equation}
Since both sets of modes are complete, the old modes can be 
expressed in terms of the new ones, i.e. 
\begin{equation}
f_{k}(\tau) = c_{+}(k) g_{k}(\tau) + c_{-}(k) g_{k}^{\ast}(\tau).
\label{match}
\end{equation}
Notice that $f_{k}(\tau)$ are normalized, as $ \tau \to - \infty$ as 
$e^{- i k\tau}/\sqrt{2k}$, since we demand that the initial state is 
the vacuum. Therefore, with this 
normalization, $c_{-}(k)$, for $\tau\to +\infty$ 
parametrizes the relevant mode-mixing.
 
Inserting Eq. (\ref{match}) back into Eq. (\ref{mdec1}) and imposing the 
continuity of the operators we also have 
\begin{eqnarray}
&& \hat{b}_{\vec{k},\alpha} = c_{+}(k) \hat{a}_{\vec{k}, \alpha} + 
c_{-}(k)^{\ast} \hat{a}_{\vec{k},\alpha}^{\dagger},
\label{bog1}\\
&& |c_{+}(k)|^2 - |c_{-}(k)|^2 =1.
\label{bog2}
\end{eqnarray}
The final value of the two-point function can then be obtained as 
\begin{equation}
\langle 0| \hat{B}_{i}(\vec{x} + \vec{r}) \hat{B}_{j}(\vec{x} |0\rangle \simeq
\frac{1}{(2\pi)^3} \int d^{3}k \biggl(\delta_{ij} -\frac{k_{i} k_{j}}{k^2}\biggr)
{\cal Q}(k) e^{- i \vec{k}\cdot \vec{r}},
\label{coprim}
\end{equation}
where now ${\cal Q}(k) \simeq  k |c_{-}(k)|^2$.
In deriving Eq. (\ref{coprim}) we used the identity
\begin{equation}
|g_{k}(\tau)|^2 =( 1 + 2 |c_{-}(k)|^2 ) |f_{k}(\tau)|^2 - c_{-}(k) c_{+}(k) 
{f_{k}^{\ast}(\tau)}^2 - c^{\ast}_{-}(k) c^{\ast}_{+}(k)
{f_{k}(\tau)}^2,
\label{osc}
\end{equation}
together with Eq. (\ref{bog2}). Notice that the the second and third terms 
in Eq. (\ref{osc}) are typically oscillating and they have been dropped 
in Eq. (\ref{coprim}).
 
Super-adiabatic amplification alone 
leads always to correlation functions falling in the general class introduced 
in Eq. (\ref{rsc}) but with ${\cal F}_{3}(r) =0$ . 
The primordial spectrum must be processed and the 
finite values of the thermal diffusivity as well as of the conductivity 
lead to an effective ultraviolet cut-off \cite{review2}.

\subsection{Primordial or astrophysical seeds?}
We are now are going to give some examples 
of magnetic fields that may be generated in the early stages of the 
life of the Universe. This topic has been already discussed in different review 
articles (see, for instance, for a short account \cite{dolgov2} and \cite{review2} for a more extended review). As 
elaborated in the introduction, the mechanisms for the generation 
of magnetic fields that may be potentially relevant for CMB physics 
can be divided, broadly speaking into two categories: mechanisms 
operating inside the Hubble radius (like phase transitions occurring, 
for instance, at the electroweak or at the QCD epoch) and mechanisms 
where the correlation scale of the field grows larger than the Hubble 
radius (as in the case of inflationary mechanisms).  Furthermore, there 
is also the appealing possibility of so-called mixed mechanisms 
where the magnetic field is created, initially, in the context of some (conventional 
or unconventional) inflationary model. Then the various modes of the 
primordial spectrum of the hypercharge field will reenter at different 
times during the radiation epoch and, at this stage, will be 
further amplified by the processes taking place inside the Hubble radius.

In the case of the electroweak phase transition (see, for instance, 
\cite{enqvist1} for a comprehensive review on this subject) the typical 
strength of the produced seeds may change depending on the order 
of the phase transition (PT). We do know that, in the framework 
of the minimal standard model of particle interaction the phase 
transition cannot be strongly first-order for values of the Higgs 
mass larger than the mass of the W boson. This conclusion was 
reached by lattice studies of the phase diagram of the electroweak 
phase transition \cite{PT1}.
Let us now consider, separately, the cases when the PT is strongly 
first-order and second order. 

In a first-order phase transition the phases of the complex order parameter of the 
nucleated bubbles are not correlated. The bubbles of the broken phase 
expand at a velocity $v \sim 10^{-3} \,c$ where $c$ is the speed 
of light.  At the epoch of the electroweak phase transition the temperature 
of the plasma is of the order of $100$ GeV while the Hubble radius, i.e. 
$H_{\rm ew}^{-1} \sim 3\, {\rm cm}$. The average distance between 
the centers of the nucleated bubbles is of the order of $10^{-8} H_{\rm ew}^{-1}$
\cite{enqvist1,aho}. The finite conductivity effects will give rise to diffusion
whose consequence will be to smooth out the oscillations of the 
created magnetic field. The magnetic flux will the  escape the intersection regions and 
penetrate the colliding bubbles where the evolution is governed by ordinary MHD.
The magnetic fields will typically have a correlation scale much smaller 
than the Hubble radius. Even granting for a rather efficient diffusion 
of the magnetic flux,  the correlation scale of the magnetic field will be at, at most, as large as the Hubble radius. In the absence of any form of inverse cascade 
we will have that the typical correlation scale at the onset of galactic rotation will 
be of the order of $10$ to $100$ A.U. (recall $ 1\, {\rm A.U.} = 1.49 \times 10^{13} 
{\rm cm}$).  The conservative  estimate of Ref. \cite{baym} is even 
smaller, i.e. the authors get a magnetic field $|\vec{B}| \sim 10^{-10}\,\,\, {\rm G}$ 
over a typical correlation scale of $0.1$ A. U. . 
As suggested in the introduction, it is plausible that some 
form of inverse cascade may eventually take place between 
the electroweak epoch and the time of electron positron annihilation
\cite{son} (see also \cite{review2} and references therein). 
If the PT is of second order the situation is even worse. The correlation 
scale  is likely to be of the order of $10/T_{\rm ew}$, i.e. 
$10^{-16} H_{\rm ew}^{-1}$ at $T_{\rm ew}$.  

There have been also ideas concerning  a possible generation 
of magnetic fields at the time of the QCD phase transition occurring 
roughly at $T\sim 140 {\rm MeV}$, i.e. at the moment when free 
quarks combine to form colorless hadrons. At this time the Hubble 
radius is of the order of $4\times 10^{4}$ m. The mechanism here 
is always related to the idea of a Biermann battery  with thermoelectric 
currents developed at the QCD time
\footnote{See later in this Section for a general discussion 
of ``batteries" in the framework of the generalized Ohm law}. Since the strange quark is heavier than the
up and down quarks there may be the possibility that the quarks develop a 
net positive charge which is compensated by the electric charge in the leptonic sector.
Again, invoking the dynamics of a first-order phase transition, it is argued that 
the shocks affect leptons and quarks in a different way so that electric currents 
are developed as the bubble wall moves in the quark-gluon plasma.  
In \cite{spergel} the magnetic field has been estimated to be $|\vec{B}|\sim {\rm G}$ 
at the time of the QCD phase transition and with 
typical scale of the order of the meter at the same epoch.

In \cite{cheng1,sigl2} it has been pointed out that, probably, 
the magnetic fields generated at the time of QCD phase transition may be 
much stronger than the ones estimated in \cite{spergel}.
The authors of \cite{cheng1,sigl2} argue that strong magnetic fields 
may be generated when the broken and symmetric phase of the theory 
coexist. The magnetic fields generated at the boundaries between quark and hadron phases can be, according to the authors, as large as $10^{6}$ G over scales of the 
order of the meter (and possibly even larger) at the time of the QCD phase transition.
At the onset of gravitational collapse and over a typical scale of the Mpc the magnetic 
field will be of the order of $10^{-26}$ G with white noise spectrum ($ m\sim 0$ 
in the notations of this section) \cite{cheng1}.

A  model of variation of gauge couplings motivated by the evolution 
of the internal dimensions \cite{maxint} leads to a variety of spectra
that may be different depending on the number of extra-dimensions 
and on their rate of evolution (see also \cite{mgvariation,bamba,orfeu}). 
While a phase of expanding 
internal dimensions may lead to large magnetic fields, the allowed 
region in the parameter space gets reduced in the case of contracting 
extra-dimensions. In \cite{maxint} only the gauge zero-modes 
have been considered.  In a recent calculation \cite{kunze},
the  r\^ole of the internal momenta has been taken into account with the 
result that also in the contracting case large magnetic seeds may be 
produced. In the case of expanding internal dimensions the spectral 
index is $m \sim -1 - n \mu $ where $n$ is the number of internal dimensions 
and $\mu$ is the exponent parametrizing, on conformal time, their evolution. 
In the case of Kasner models \cite{maxint} $\mu = \sqrt{3n/(n + 2)}$. If internal 
dimensions contract we have, effectively, $m \sim 0$  \cite{kunze}.

Large-scale magnetic fields can also be produced in the context of inflationary 
models. Here the idea is that quantum fluctuations of an  Abelian gauge field (for instance the hypercharge field) are inside the Hubble radius during inflation. 
Then, these fluctuations exit the horizon and their correlation scale grows faster 
than the Hubble radius itself. The amount of amplification achieved in this 
processes depends crucially on the amount of breaking 
of conformal invariance of the evolution equation of the Abelian gauge fields. 
A lot of ideas concerning the possible 
breaking of conformal invariance (see, for instance, \cite{review2} and 
\cite{dolg2}) 
have been proposed so far and it would be impossible 
to summarize them all in this topical review. 

Instead going through the various ideas it may be more useful to give 
further details about a specific way of breaking conformal invariance
like the one suggested in the context of string cosmological models \cite{maxst}.
The string effective action contains a direct coupling between the 
(scalar) dilaton field and the electromagnetic field that appears 
even at tree level in the loop expansion and to lowest 
order in the string tension expansion. To lowest order the 
action will then be 
\begin{equation}
S = -\frac{1}{4} \int d^{4} x \sqrt{-g}  e^{-\varphi} Y_{\alpha\beta}
Y^{\alpha\beta},
\end{equation}
where $Y_{\alpha\beta}$ is the field strength. If the gauge coupling evolves 
during the inflationary dynamics, then the magnetic fields will be 
amplified over different length-scales. In particular, in the model 
presented in \cite{maxst}, the gauge coupling, i.e. 
$e^{\varphi/2}$ is a growing function of the 
cosmic time coordinate. When the Universe becomes 
dominated by radiation (after inflation) the gauge coupling freezes 
to a value that is of order of $10^{-2}$.  More specifically, there 
are two phases that characterize the model. In the first phase 
the gauge coupling increases sharply (the so-called dilaton driven phase).
In the second phase the gauge coupling is just slowly increasing (the so-called stringy phase).

In the context of string cosmological models there are two possible ranges in the values of $m$, i.e. the spectral index of the produced magnetic fields. If the 
relevant modes (i.e. comoving scales 
larger than the Mpc scale) exit the Hubble radius for the first 
time during the so-called dilaton driven phase, the resulting 
spectrum leads to an $m$  that could be written as \cite{maxst}
\begin{equation}
m = - \frac{\sqrt{3} + 3 \sqrt{1 - \Sigma}}{\sqrt{3} + \sqrt{ 1 - \Sigma}},
\label{mst}
\end{equation}
where $\Sigma = \sum_{i} \beta_{i}^2$ represents the possible 
effect of internal dimensions whose radii $b_{i}$ shrink, during 
the pre-big bang phase, like $(-t)^{\beta_{i}}$ for $t\to 0^{-}$.
In the case $\Sigma = 0$ we will have $m \sim - \sqrt{3}$.

The second possibility is that the relevant large-scale modes 
left the Hubble radius during the so-called string phase. During 
this phase the ratio between the (cosmic) time derivative 
of the dilaton field and the Hubble rate is roughly constant.
Defining as $\beta$ the mentioned ratio $\dot{\varphi}/2 H$, 
we will have that $m \simeq 1 -2 \beta$. If the value of $\beta$ gets 
frozen at the value corresponding to the last e-folds of the 
pre-big bang phase we will have, for instance, allowed values between 
$\beta \sim 1.8$ and $\beta\sim2$ 
corresponding to $m \simeq -3$ (i.e. $-2.9$, $-2.8$). This 
parameter can be computed once a specific model of the 
high-curvature (stringy) regime is given but it cannot be obtained, at 
present, from more general arguments. Notice that, in this case, 
$B_{L}$ may even be of the order of $10^{-10}$ corresponding 
to a ratio between the magnetic and radiation energy density 
of the order of $10^{-8}$ \cite{maxst}.

It is also possible to generate large scale magnetic fields within 
the Abelian Higgs model where the Higgs current densities 
allow to obtain a (minute) amplification of the hypermagnetic field.
The amplitude of these fields is irrelevant for applications (i.e. the produced 
field intensity is thirty orders of magnitude smaller than the 
most optimistic dynamo requirements). The typical spectral slope 
is $m = 1$ that corresponds to the magnetic spectrum of the vacuum 
once the canonical quantum mechanical normalization is imposed 
on the vector potentials. Slightly different spectra can be obtained 
in the case of the models proposed in \cite{hec1,hec2,hec3} where 
various charged scalars are considered together (see also \cite{kari}).

It is appropriate to comment here 
on the possibility that large-scale magnetization is a purely 
astrophysical phenomenon.  As in the case of the primordial 
hypothesis, it is rather difficult to summarize, in a reasonably self-contained 
manner all the hypothesis that have been put forward in the last fifty years.
An (incomplete) list of proposals can be found, for instance 
in \cite{review2,laz1,rees1}.  The modest aim 
of the remaining part of this section will therefore 
be to introduce the general framework. 

Many (if not all) the astrophysical mechanisms proposed so far
are related to what is called, in the jargon, a battery. 
The idea is, in short the following.  MHD, as previously 
stressed in this section, can be ``derived" from a two-fluid 
theory by appropriately defined one-fluid variables that 
obey a set of effective Maxwell equations supplemented 
by the Ohm law (i.e. Eq. (\ref{ohmic})) and coupled to the 
Navier-Stokes equation. Now, we already remarked right before 
Eq. (\ref{ohmic}) that the proposed form of the Ohmic electric 
field would contain, in principle, also a thermoelectric 
term that is a remnant of the subtraction of the momentum 
conservation equations that hold separately (for electrons and ions) 
in the two-fluid description from which MHD is derived.
The explicit form of the generalized Ohmic electric field 
in the presence of thermoelectric corrections can be written 
as \footnote{For simplicity, we shall neglect 
the Hall contribution arising in the generalized Ohm law. The Hall 
contribution would produce, in Eq. (\ref{thermo}) a term $\vec{J}\times \vec{B}/
n_{\rm e} e$ that is of higher order in the magnetic field and that is 
proportional to the Lorentz force. The Hall term will play no r\^ole 
in the subsequent considerations. However, it should be borne in mind 
that the Hall contribution may be rather interesting in connection 
with the presence of strong magnetic fields like the ones 
of neutron stars (i.e. $10^{13}$ G). This occurrence is even more 
interesting since in the outer regions of neutron stars strong density 
gradients are expected.}
\begin{equation}
\vec{E} = - \vec{v} \times \vec{B} + 
\frac{\vec{\nabla}\times \vec{B}}{4\pi \sigma} - 
\frac{\vec{\nabla}P_{\rm e}}{ e n_{\rm e}}.
\label{thermo}
\end{equation}
By comparing Eq. (\ref{ohmic}) with Eq. (\ref{thermo}), it is clear 
that the additional term at the right hand side, receives contribution
from a temperature gradient. In fact, restoring for a moment the Boltzmann
constant $k_{B}$ we have that since  $P_{\rm e} = k_{B} \, n_{\rm e} \, T_{\rm e}$, the additional term depends upon the gradients of the temperature, hence 
the name thermoelectric. It is interesting to see under which conditions 
the curl of the electric field receives contribution from the thermoelectric 
effect. Taking the curl of both sides of Eq. (\ref{thermo}) 
we obtain 
\begin{equation}
\vec{\nabla}\times \vec{E} = \frac{1}{4\pi \sigma} \nabla^2 \vec{B} + 
\vec{\nabla}(\vec{v} \times \vec{B}) - \frac{\vec{\nabla}n_{\rm e} \times 
\vec{\nabla} P_{\rm e}}{e n_{\rm e}^2}  = - \frac{\partial B}{\partial t},
\label{diffusivity}
\end{equation}
where the second equality is a consequence of Maxwell's equations.
From Eq. (\ref{diffusivity}) it is clear that the evolution of the magnetic field 
inherits a source term iff (i.e. if and only if) the gradients in the pressure and electron 
density are not parallel. If $\vec{\nabla} P_{\rm e} \parallel \vec{\nabla} n_{\rm e}$ a fully valid solution of Eq. (\ref{diffusivity}) is $\vec{B}=0$. In the opposite 
case a seed magnetic field is naturally provided by the thermoelectric  term.
The usual (and rather general) observation that one can make in connection with the geometrical properties of the thermoelectric term
is that cosmic ionization fronts may play an important r\^ole. 
For instance, when quasars emit ultraviolet photons, cosmic 
ionization fronts are produced. Then the intergalactic medium 
may be ionized. It should also be recalled, however, that the temperature 
gradients are usually normal to the ionization front. In spite of this, it is 
also plausible to think that density gradients can arise in arbitrary 
directions due to the stochastic nature of density fluctuations.

In one way or in another, astrophysical mechanisms for the generation 
of magnetic fields use an incarnation of the thermoelectric effect. 
In the sixties and seventies, for instance, it was rather popular to think 
that the correct ``geometrical" properties of the thermoelectric term
may be provided by a large-scale vorticity. As it will also be discussed 
later, this assumption seems to be, at least naively, in contradiction
with the formulation of inflationary models whose prediction would actually 
be that the large-scale vector modes are completely washed-out by 
the expansion of the Universe. Indeed, all along the eighties and nineties 
the idea of primordial vorticity received just a minor attention. 

The attention then focused on the possibility that objects of rather 
small size may provide intense seeds. After all we do know that 
these objects may exist. For instance the Crab nebula has 
a typical size of a roughly 1 pc and a magnetic field that is a 
fraction of the m G. These seeds will then combine and diffuse 
leading, ultimately, to a weaker seed but with large correlation scale.
This aspect, may be, physically, a bit controversial since we do
observe magnetic fields in galaxies and clusters that are ordered over very 
large length scales. It would then seem necessary  that the seed fields 
produced in a small object (or in several small  objects) undergo 
some type of dynamical self-organization whose final effect
is a seed coherent over length-scales 4 or 5 orders of magnitude 
larger than the correlation scale of the original battery.

An interesting idea could be that qualitatively different 
batteries lead to some type of conspiracy that may produce a strong 
large scale seed. In \cite{rees1} it has been suggested that Population III
stars may become magnetized thanks to a battery operating at 
stellar scale. Then if these stars would explode as supernovae (or if they 
would eject a magnetized stellar wind) the pre-galactic environment 
may be magnetized and the remnants of the process incorporated 
in the galactic disc. In a complementary perspective, a similar 
chain of events may take place over a different physical scale. A battery 
could arise, in fact in active galactic nuclei at high red-shift. Then 
the magnetic field could be ejected leading to intense fields in the lobes 
of ``young" radio-galaxies. These fields will be somehow inherited by 
the ``older" disc galaxies and the final seed field may be, according 
to \cite{rees1} as large as $10^{-9}$ G at the pre-galactic stage.

Up to now only a necessarily reduced account of the primordial and of the 
astrophysical hypotheses has been presented. In spite of the mechanism 
leading  to a seed field, the crucial problem will be how 
to go from the seed to the actual galactic (or even cluster) magnetic field.
The galactic rotation period is of the order of $3\times 10^{8}$ yrs. This 
scale should be compared with the typical age of the galaxy. All along this  rather large dynamical time-scale the effort has been directed, from the fifties,
to the justification that a substantial portion of the kinetic energy of the system 
(provided by the differential rotation) may be converted into magnetic 
energy amplifying, in this way, the seed field up to the observed value 
of the magnetic field, for instance in galaxies and in clusters. 
In recent years a lot of progress has been made both in the context 
of the small and large-scale dynamos. This progress was also driven 
by the higher resolution of the numerical simulations and by the 
improvement in the understanding of the largest magnetized 
system that is rather close to us, i.e. the sun. More complete 
accounts of this progress can be found in the second 
paper of Ref. \cite{lazarian} and, 
more comprehensively, in Ref. \cite{brandenburg}. The main 
ingredients of the large-scale dynamo can be also found in \cite{review2}.
Apart from the aspects involving solar physics and numerical 
analysis, better physical understanding of the r\^ole of the 
magnetic helicity in the dynamo action has been reached. This 
point is crucially connected with the two conservation laws 
arising in MHD, i.e. the magnetic flux and magnetic helicity 
conservations whose relevance has been already emphasized, 
respectively, in Eqs. (\ref{flux}) and (\ref{hel}).
Even if the rich interplay between small and large scale dynamos 
is rather important, let us focus on the problem of large-scale 
dynamo action that is, at least superficially, more central for the 
considerations developed in the present paper. 
In short, the picture of large-scale dynamos can be summarized 
as follows. In a turbulent environment the MHD equations 
can be appropriately averaged with the result that the mean magnetic seed 
is amplified exponentially fast. This approach  is 
described in detail in \cite{brandenburg} (see also \cite{kulsrud1}).
The most important point to bear in mind is that the turbulent 
velocity field (whose correlation scale is typically shorter 
than the  correlation scale of the magnetic seed) {\em must} lack 
mirror symmetry. In other words the velocity field must be 
parity-odd. This occurrence necessarily implies that the 
(averaged) magnetic diffusivity equation inherits a 
term that is proportional to the curl of the magnetic field 
though a coefficient, the celebrated $\alpha$ term, that would vanish 
if the velocity field would be parity-even. It can be demonstrated 
that the $\alpha$ term is indeed proportional, in turn, to the average 
{\em kinetic} helicity, i.e. $\langle \vec{v}\cdot\vec{\nabla} \times 
\vec{v}\rangle$. The $\alpha$ term decreases more slowly, at large 
length-scales, than the diffusivity terms. Therefore, for sufficiently 
large scales the magnetic seed can be efficiently (exponentially?) 
amplified. For long time this 
oversimplified picture was thought to be essentially consistent 
even if computationally challenging. 

Already at a qualitative level it is clear that there is a clash between 
the absence of mirror-symmetry of the plasma, the quasi-exponential 
amplification of the seed and the conservation of magnetic flux and 
helicity in the high (or more precisely infinite) conductivity limit. 
The easiest clash to understand, intuitively, is the flux 
conservation versus the exponential amplification: both flux freezing 
and exponential amplification have to take place in the 
{\em same} superconductive (i.e. $\sigma^{-1} \to 0$) limit. 
The clash between helicity conservation and dynamo action 
can be also understood in general terms: the dynamo action implies 
a topology change of the configuration
since  the magnetic flux lines cross each other constantly \cite{lazarian}.

One of the recent progress in this framework  is a more consistent 
formulation of the large-scale dynamo problem\cite{lazarian,brandenburg}: 
 large scale dynamos produces small scale helical fields that quench 
 (i.e. prematurely saturate) the $\alpha$ effect.  In other words, the 
 conservation of the magnetic helicity can be seen, according to the 
 recent view, as a fundamental constraint on the dynamo action.
 In connection with the last point, it should be mentioned that, in the past, 
 a rather different argument was suggested \cite{kulsrud2}: it was argued that 
 the dynamo action not only leads to the amplification of the large-scale field 
 but also of the random field component. The random field would then 
 suppress strongly the dynamo action. According to the considerations 
 based on the conservation of the magnetic helicity this argument seems 
 to be incorrect since the increase of the random component would 
 also entail and increase of the rate of the topology change, i.e. 
 a magnetic helicity non-conservation.
 
 In summary we can therefore say that:
 \begin{itemize}
 \item{} both the primordial and the astrophysical hypothesis 
 for the origin of the seeds demand an efficient (large-scale) 
 dynamo action;
 \item{} due to the constraints arising from the conservation 
 of magnetic helicity and magnetic flux the values of the 
 required seed fields may turn out to be larger than previously 
 thought at least in the case when the amplification is only driven 
 by a large-scale dynamo action \footnote{The situation may change if 
 the magnetic fields originate from the combined action  of small
 and large scale dynamos like in the case of the two-step 
 process described in \cite{rees1}.};
 \item{} magnetic flux conservation during gravitational collapse 
 of the protogalaxy 
 may increase, by compressional amplification, the initial seed 
 of even 4 orders of magnitude;
 \item{} compressional amplification, as well as large-scale dynamo, 
 are much less effective in clusters: therefore, the magnetic field
 of clusters is probably connected to the specific way the dynamo saturates, and, in this sense, harder to predict from a specific value of the initial 
 seed.
 \end{itemize}

 \renewcommand{\theequation}{5.\arabic{equation}}
\section{Magnetized metric perturbations}
\setcounter{equation}{0}
Consider a conformally flat background geometry 
\begin{equation}
\overline{g}_{\mu\nu} = a^2(\tau) \eta_{\mu\nu}
\end{equation}
where $\eta_{\mu\nu} = {\rm diag}(1,\, -1,\, -1,\,-1)$ is the flat Minkowski 
metric. Then 
the fluctuations of the metric may be written as 
\begin{equation}
\delta g_{\mu\nu} = \delta_{\rm t} g_{\mu\nu} + \delta_{\rm v} g_{\mu\nu} + 
\delta_{\rm s} g_{\mu\nu},
\label{flucg}
\end{equation}
where $\delta_{\rm t}$, $\delta_{\rm v}$ and $\delta_{\rm s}$ 
denote the tensor, vector and scalar fluctuation of the geometry.

The evolution  of the various modes of the geometry 
are sourced by the corresponding modes of the energy-momentum
tensor of the fluid sources and of the magnetic field.  In particular, 
denoting with ${\cal T}_{\mu\nu}$ the energy-momentum 
tensor of the magnetic field we will have, following the notation of 
Eq. (\ref{flucg}), that 
\begin{equation}
\delta {\cal T}_{\mu\nu} = \delta_{\rm t} {\cal T}_{\mu\nu} + \delta_{\rm v} {\cal T}_{\mu\nu} + \delta_{\rm s} {\cal T}_{\mu\nu}.
\label{flucT}
\end{equation}
The fluctuations $\delta {\cal T}_{\mu\nu}$ 
are {\em quadratic} in the intensity of the fully inhomogeneous 
magnetic fields. This means, that a non-Gaussian signal should be, in general, 
expected. The rationale for this statement is that while  
the fluctuations of  fully inhomogeneous magnetic fields may be  Gaussian
(see, for instance, Eq. (\ref{stoch1})), 
the fluctuations of the energy-momentum tensor will 
be distributed as the square of a Gaussian variable and will therefore 
be non-Gaussian. Recently, more accurate calculations of the possible 
non-Gaussian effects arising in the presence of fully inhomogeneous 
magnetic fields have been presented \cite{robert}.   

In the following the evolution equations of tensor, vector and scalar modes 
will be treated within a (first-order) ``Bardeen" approach \cite{bardeen} 
(see also \cite{cmbrev} for an introduction to the key features 
of the problem in the absence of magnetic fields).
The evolution of fully inhomogeneous magnetic fields can  
be usefully treated also within a covariant formalism. For this 
type of analysis see the interesting studies reported in Refs. \cite{tsagas1,tsagas2,tsagas3}.

\subsection{Magnetized tensor modes}
The tensor fluctuations of the metric defined as 
\begin{equation}
\delta_{\rm t} g_{ij} = - a^2 h_{ij},\qquad \partial_{i} h^{i}_{j} =0,\qquad 
h_{i}^{i} =0,
\label{tensdef}
\end{equation}
are automatically invariant under infinitesimal coordinate transformations.
Equation (\ref{tensdef}) implies that $h_{ij}$ carries two 
independent degrees of freedom.
From the fluctuation of the Ricci tensor it is easy to obtain that 
the evolution equations are 
\begin{equation}
{h_{i}^{j}}'' + 2 {\cal H} {h_{i}^{j}}' - \nabla^2 h_{i}^{j} = - 16\pi G a^2 
\delta_{\rm t} {\cal T}_{i}^{j},
\label{tenseq}
\end{equation}
where, as mentioned before, $\delta_{\rm t} {\cal T}_{i}^{j}$ is the transverse 
and traceless component of the energy-momentum 
tensor of the electromagnetic fields. Given the specific 
magnetic field configuration, Eq. (\ref{tenseq}) allows to determine 
the amplitude of the gravitational waves.
The determination of $h_{ij}$ allows the determination 
of the tensor  Sachs-Wolfe contribution, i.e.
\begin{equation}
\biggl(\frac{\Delta T}{T}\biggr)_{\rm t} = - 
\frac{1}{2} \int_{\tau_i}^{\tau_{f}} h_{ij}' n^{i} n^{j} d\tau. 
\label{TSW}
\end{equation}
In the tensor case the fluctuations of the geometry do not mix with the fluctuations 
of the fluid sources. Furthermore, in the tensor case, as it is well known, only 
the Sachs-Wolfe integral contributes. Tensor modes have been discussed
in a number of papers. In \cite{q5} and \cite{mgknot} the 
possible stochastic backgrounds of gravitational radiation stemming 
from inhomogeneous magnetic fields have been computed with 
particular attention to configurations carrying non-vanishing 
magnetic gyrotropy. Recently gravitational waves produced through 
a phase of helical turbulence have been discussed in \cite{tk4}.  For a 
discussion of the interplay of gravitational waves and inhomogeneous 
magnetic fields see also \cite{tsagas4}.

\subsection{Magnetized vector modes}
The evolution equations arising in the case of vector modes are a bit more 
complicated than in the case of tensor modes. The vector modes of the 
geometry are parametrized in the following way 
\begin{equation}
\delta_{\rm v} g_{0 i} = - a^2 Q_{i},\qquad \delta_{\rm v} g_{i j} = 
a^2 ( \partial_{i} W_{j} + \partial_{j} W_{i}),
\label{vectdef}
\end{equation}
where $\partial_{i} Q^{i}=0$ and $\partial_{i}W^{i} =0$.

The perturbed Einstein and covariant conservation equations read 
\begin{eqnarray}
&& \delta_{\rm v} R_{\mu}^{\nu} = 8\pi G \bigl( \delta_{\rm v} 
T_{\mu}^{\nu} + \delta_{\rm v} {\cal T}_{\mu}^{\nu} \bigr),
\label{einV}\\
&& \nabla_{\mu} \delta_{\rm v} T^{\mu \nu} + 
\nabla_{\mu} \delta_{\rm v} {\cal T}^{\mu\nu} =0
\label{covV},
\end{eqnarray}
where  
\begin{eqnarray}
&& \delta_{\rm v} T_{\mu}^{\nu} = ( p + \rho) (u_{\mu} \delta_{\rm v} u^{\nu} + 
\delta_{\rm v} u_{\mu} u^{\nu}), 
\label{sourceV}\\
&& \delta_{\rm v} {\cal T}_{\mu}^{\nu} = 
\frac{1}{4\pi} \biggl( - F_{\mu\alpha} F^{\nu\alpha} + \frac{1}{4} 
\delta_{\mu}^{\nu} F_{\alpha\beta}F^{\alpha\beta} \biggr).
\label{MXV}
\end{eqnarray}
Equations (\ref{einV}) and (\ref{MXV}) have to be supplemented by 
 Maxwell's equations together with the generalized Ohm's law
\begin{eqnarray}
&& \nabla_{\mu} F^{\mu\nu} = 4 \pi j^{\mu}
\nonumber\\
&& j^{\mu} = \sigma_{\rm c} F^{\mu\nu} u_{\nu}
\end{eqnarray}
Recall first that
\begin{eqnarray}
&& \delta_{\rm v} R_{0}^{i} = \frac{2}{a^2} Q^{i} ( {\cal H}'  - {\cal H}^2)
+ \frac{1}{2 a^2} \nabla^2 V^{i},
\nonumber\\
&& \delta_{\rm v} R_{i}^{j} = \frac{1}{ 2 \, a^2} \bigl[ (\partial_{i} V^{j} 
+ \partial^{j} V_{i})' + 2 {\cal H} (\partial_{i} V^{j} +\partial^{j} V_{i}) \bigr].
\end{eqnarray}
where
\begin{equation}
\vec{V} = \vec{Q} + \vec{W}'.
\end{equation}
The vector $\vec{V}$ is gauge-invariant. In fact, for infinitesimal 
gauge transformations preserving the vector nature of the fluctuation we
have, indeed, 
\begin{eqnarray}
&& Q_{i} \to \tilde{Q}_{i} = Q_{i} - \zeta_{i}',
\label{Qgauge}\\
&& W_{i} \to \tilde{W}_{i}= W_{i} + \zeta_{i}.
\label{Wgauge}\\
\end{eqnarray}

From Eq. (\ref{sourceV}), recalling that
\begin{equation}
u_{0} \delta_{\rm v} u^{i} = -( {\cal V}^{i} + Q^{i}) ,
\end{equation}
we simply have 
\begin{equation}
\delta_{\rm v} T_{0}^{i} = -(p + \rho) ({\cal V}^{i} + Q^{i}).
\end{equation}
Equation (\ref{MXV}) become, in explicit terms,
\begin{eqnarray}
&& \delta_{\rm v} {\cal T}_{0}^{i} = \frac{1}{4\pi} (\vec{{\cal E}} \times \vec{{\cal B}})^{i}  \equiv \frac{1}{4\pi a^4} (\vec{ E} \times \vec{ B})^{i},
\label{mx0iV}\\
&& \delta_{\rm v} {\cal T}_{i}^{j} = \frac{1}{4\pi} \bigl[ {\cal E}_{i} {\cal E}^{j} 
+ {\cal B}_{i} {\cal B}^{j} - \frac{1}{2} \delta_{i}^{j} \bigl( \vec{{\cal B}}^2 + 
\vec{{\cal E}}^2\bigr) \bigr] 
\nonumber\\
&&\equiv \frac{1}{4\pi a^4} \bigl[ E_{i} E^{j} 
+  B_{i}  B^{j} - \frac{1}{2} \delta_{i}^{j} \bigl( \vec{ B}^2 + 
\vec{ E}^2\bigr) \bigr],
\label{mxijV}
\end{eqnarray}
where the following conventions have been adopted
\begin{eqnarray}
&& F_{0i} = a^2 {\cal E}_{i},\,\,\,\,\,\,\,\,\,\,\,\,\, F^{0i} = - \frac{1}{a^2} {\cal E}^{i},
\nonumber\\
&& F_{ij} = - a^2 \epsilon_{ij k} {\cal B}^{k},\,\,\,\,\,\,\,\,\,\,\,\,\,
F^{ij} = -\frac{1}{a^2} \epsilon^{ijk} {\cal B}_{k},
\nonumber\\
&& \vec{E} = a^2 \vec{{\cal E}}, \,\,\,\,\,\,\,\,\,\,\,\,\, \vec{B} = a^2 
\vec{{\cal B}}.
\end{eqnarray}

Therefore, the perturbed set of equations can be written, in the gauge 
$W_{i} =0$, 
\begin{eqnarray}
&&\nabla^2 \vec{Q} = - 16\pi G a^2 ( p + \rho) \vec{{\cal V}} + \frac{16 \pi G}{\sigma a^2} \vec{F}_{B}(\vec{x}),
\label{V1}\\
&& \nabla^2( \vec{Q}' + 2 {\cal H} \vec{Q}) = \frac{16\pi G}{a^2} \vec{F}_{B}(\vec{x}),
\label{V2}\\
&& \vec{{\cal V}}'  + \biggl[ 4  {\cal H} + \frac{p' + \rho'}{p+ \rho}\biggr]
\vec{ {\cal V}} + \frac{\vec{F}_{B}(\vec{x})}{ a^4 (p +\rho)} =0,
\label{V3}
\end{eqnarray}
where
\begin{equation}
\vec{F}_{B} = \frac{1}{4\pi}( \vec{\nabla}\times \vec{B})\times \vec{B}.
\end{equation}
Notice that the divergence-full part of the Lorentz force (i.e. $\vec{\nabla}\cdot 
\vec{F}_{B} $) does not contribute to the evolution equations of the 
vectors but it does contribute to the evolution equations of the scalar modes.
The second term at the right hand side of Eq. (\ref{V1}) contains, in the denominator, $\sigma = a \sigma_{\rm c}$, i.e. the rescaled conductivity. This term, coming from the MHD form of the Poynting vector, is negligible 
in the infinite conductivity limit and it is usually dropped.

Formally, the solution of the above system of equations can be written as 
\begin{eqnarray}
&&\vec{V}(\vec{x},\eta) = - \frac{1}{a^4(\eta) (p+\rho)} \int \vec{F}_{B}
(\vec{x},\eta) d\eta - \frac{\vec{C}(\vec{x})}{16\pi G a^4(\eta) (p+ \rho)},
\nonumber\\
&&\nabla^2 \vec{Q} = \frac{16\pi G}{a^2(\eta)} \int \vec{F}_{B}(\vec{x},\eta) d\eta 
+ \frac{\vec{C}(\vec{x})}{a^2(\eta)}.
\end{eqnarray}
Finally, the vector contribution to the Sachs-Wolfe effect can be written, in the gauge 
$W_{i} =0$ as 
\begin{equation}
\biggl(\frac{\Delta T}{T}\biggr)_{\rm v} = [ -\vec{{\cal V}}\cdot \vec{n}]_{\tau _{i}}^{\tau_{f}} + \frac{1}{2} \int_{\tau_{i}}^{\tau_{f}}
(\partial_{i} Q_{j} + \partial_{j} Q_{i}) n^{i} n^{j} d\tau.
\end{equation}

In the absence of large-scale magnetic field, the only solution of the above system is the one denoted by the integration constant $C$. In this 
case, if the Universe expands, the vector modes can only decay 
(see, however, \cite{bat,mgvect1,mgvect2}).  

In a series of papers Barrow and Subramanian \cite{jkan1,jkan2}
analyzed the effects of tangled magnetic fields both on the 
temperature and polarization anisotropies. 
More recently the vector
modes have been analyzed numerically in Ref. \cite{lewis1} (see also 
\cite{lewis2}).  The obtained results are in qualitative agreement with 
the semi-analytical estimates presented in \cite{jkan1,jkan2}.
 In \cite{lewis1} the effect of the neutrinos has been 
accurately parametrized in the evolution equations. In the equations 
for the vector modes written above, the (divergenceless) velocity 
field $\vec{\cal V}$ has been taken generic. However, when the plasma 
content before decoupling is specified, $\vec{{\cal V}}$, i.e. the total 
velocity field will receive contribution from all the species present in the 
plasma. A similar situation will be encountered in the case of scalar 
modes and will be briefly discussed there.

\subsection{Magnetized scalar modes}

The evolution of scalar fluctuations is, technically, more complicated than the 
vector and tensor problems. In the case of scalar fluctuations, the magnetic fields 
enter the evolution equations through the magnetic energy density 
and through the divergence of the Lorentz force. 
Furthermore, in the scalar case, the possible gauge choices 
increase with respect to the case of the vector modes. 

In the absence of magnetic fields, we do know that the most general 
initial conditions for CMB anisotropies consist of one adiabatic mode 
and four non-adiabatic (or isocurvature) modes (see, for 
instance, \cite{cmbrev}).  In the case of adiabatic modes 
the fluctuations in the specific entropy vanish for typical length-scales 
larger than the Hubble radius. This occurrence implies a specific 
relation among the density contrasts  of the various 
species present in the plasma. 
Consider, indeed, the simple case of a system formed by cold dark matter (CDM)
and radiation. In this case, the specific entropy $\varsigma$
 and its relative fluctuations $ {\cal S}$ can be written as 
 \begin{equation}
 \varsigma= \frac{T^3}{n_{\rm c}},\qquad {\cal S} = \frac{\delta\varsigma}{\varsigma} = 
 \frac{3}{4} \delta_{\gamma} - \delta_{\rm c}.
 \label{ad1}
 \end{equation}
 where $n_{\rm c}$ is the number density of CDM particles; $\delta_{\gamma}$ and 
 $\delta_{\rm c}$ are the density contrasts of the fluctuations in radiation and in CDM.
 If ${\cal S} \simeq 0$ for scales larger than the Hubble radius, the fluctuations 
 are said to be adiabatic. 
 
 This naive definition can be generalized in various ways. One can, for instance, give
 a more general gauge-invaraint definition by introducing two generic fluids, the 
 a-fluid and the b-fluid. In this case it is rather natural to define 
 the two gauge-invariant variables 
 \begin{eqnarray}
 && \zeta_{\rm a} = - \psi - {\cal H} \frac{\delta\rho_{\rm a}}{\rho_{\rm a}'},
 \label{adensity}\\
 && \zeta_{\rm b} = - \psi - {\cal H} \frac{\delta\rho_{\rm b}}{\rho_{\rm b}'},
 \label{bdensity}
 \end{eqnarray}
that can be interpreted, respectively,  as the curvature fluctuations on uniform 
a-density and b-density hypersurfaces (see, for instance, \cite{mgentr}  and references therein). The (total) curvature 
fluctuations on uniform density hypersurfaces are then 
\begin{equation}
\zeta = \frac{\rho_{\rm a}'}{\rho'} \zeta_{\rm a} + \frac{\rho_{\rm b}'}{\rho'} \zeta_{\rm b}
\equiv  - \psi - {\cal H} \frac{\delta\rho}{\rho'},
\label{tot} 
\end{equation}
where $\rho = \rho_{\rm a} + \rho_{\rm b}$. From Eqs.  (\ref{adensity}) and 
(\ref{bdensity}) it is possible to define 
the generalized entropy fluctuations as 
\begin{equation}
{\cal S} = - 3 ( \zeta_{\rm b }- \zeta_{\rm a}) = \frac{\delta_{\rm b}}{1 + w_{\rm b}} - 
\frac{\delta_{\rm a}}{1 + w_{\rm a}}, 
\label{entrgen}
\end{equation}
where the second equality follows from the first equality by using the covariant 
conservation equation for the each of the two fluids; $w_{\rm a}$ and $w_{\rm b}$ 
are, respectively, the barotropic indices for the a-fluid and the b-fluid. 
In the case 
$w_{\rm b} = w_{\gamma} = 1/3$ and $w_{\rm a}=0$  Eq. (\ref{entrgen}) 
reproduces Eq. (\ref{ad1}).

If we now consider the plasma after equality but before decoupling, 
the total entropy fluctuations will vanish iff 
\begin{equation}
\delta_{\gamma} \simeq \delta_{\nu} \simeq \frac{4}{3} \delta_{\rm c} \simeq 
\frac{4}{3} \delta_{\rm b}
\label{ad2}
\end{equation}
where the density contrast in the neutrinos and in the baryons, i.e. 
$\delta_{\nu}$ and $ \delta_{\rm b}$, have been introduced.

If magnetic fields are included in the game, then 
the adiabatic mode, together with the four isocurvature modes, can be generalized. 
The logic will then be the following:
\begin{itemize}
\item{} the evolution equations of the whole system must be written including 
the contribution of the large-scale magnetic fields;
\item{} the system includes, on top of the magnetic field and of the 
scalar fluctuations of the geometry also the neutrinos, the photons, the 
baryons and the CDM particles;
\item{} the equations must be solved when the relevant modes are 
larger than the Hubble radius;
\item{} the obtained solutions will serve as initial conditions 
for the lowest multipoles of the Boltzmann hierarchy.
\end{itemize}

It is difficult, in the scalar case, to go trough all the previous steps 
without changing gauge. Indeed it is well known, also in the absence 
of large-scale magnetic fields, that some isocurvature modes 
(for which the presence of anisotropic stresses is crucial) are regular 
in the synchronous gauge but are divergent (at early times) 
in the longitudinal gauge. This is just an example and, in the following,
 the magnetized adiabatic mode will be mainly discussed, quoting, when 
 appropriate, some results of more  technical investigations 
 \cite{magnetizedincon}.

\subsubsection{Magnetized adiabatic mode}

Consider, for simplicity, the system of scalar fluctuations in the longitudinal gauge
where the metric fluctuations can be written in terms of the two functions $ \phi$ and $\psi$, i.e. 
\begin{equation}
\delta_{\rm s} g_{00} = 2 a^2 \phi,\,\,\,\,\,\,\,\,\,\,\,\,\,\, 
\delta_{\rm s} g_{i j} = 2 a^2 \psi \delta_{i j}.
\label{longgauge}
\end{equation}
In the following, the relevant equations of the system will be discussed 
starting with the evolution equations of the fluctuations. As 
mentioned before, the plasma content is assumed to be formed 
by neutrinos, photons, baryons, CDM particles and fully 
inhomogeneous maggnetic fields. 

In the gauge of Eq. (\ref{longgauge}),  the Hamiltonian constraint reads 
\begin{equation}
- 3 {\cal H} ( {\cal H} \phi + \psi') - k^2 \psi = \frac{3}{2} {\cal H}^2 [ ( R_{\nu} \delta_{\nu} + ( 1 - R_{\nu}) \delta_{\gamma}) + \Omega_{\rm B}(k) + 
\Omega_{\rm b} \delta_{\rm b} + \Omega_{\rm c} \delta_{\rm c}],
\label{p00lex}
\end{equation}
where, for $N_{\nu}$ species of massless neutrinos,   
\begin{equation}
R= \frac{7}{8} N_{\nu} \biggl( \frac{4}{11}\biggr)^{4/3},\,\,\,\,\,\,\, R_{\nu}  = \frac{R}{1 + R},\,\,\,\,\,\,\,\, R_{\gamma} = 1 - R_{\nu},
\end{equation}
so that $R_{\nu}$ and $R_{\gamma}$ represent the fractional contributions of photons and neutrinos to the total density at early times 
deep within the radiation-dominated epoch.

In Eq. (\ref{p00lex}) the contribution of the magnetic energy density has been parametrized, in Fourier space, as  
\begin{eqnarray}
\Omega_{\rm B}(k,\eta) = \frac{\rho_{\rm B}}{\rho} =\frac{1}{8\pi \rho a^4} \int d^{3} p\,\, B_{i}(|\vec{p} - \vec{k}|)\, B^{i}(p).
\label{OMB}
\end{eqnarray}
The appearance of convolutions in the expressions  of the magnetic energy density is a direct consequence 
of the absence of a uniform component of the magnetic field whose background contribution, as 
repeatedly stressed, is vanishing. 
Notice also  that $\rho$ appearing in Eq. (\ref{OMB})  is the energy density of the {\em dominant} 
component of the fluid, so, for instance, in the radiation-dominated epoch 
$\rho\equiv \rho_{\gamma}$, $a^4 \rho_{\gamma} \sim {\rm constant}$ and $\Omega_{B}\simeq {\rm constant}$. 

While the Hamiltonian constraint relates the metric fluctuations (and their 
first time derivatives) to the density contrasts of the various species, 
the momentum constraints relates the metric fluctuations (and their 
first derivatives) to the divergence-full part of the peculiar velocity 
field and to the MHD expression of the Poynting vector\footnote{In the 
following $\theta_{\gamma} = \partial_{i} v^{i}_{\gamma} $, 
$\theta_{\rm c} = \partial_{i} v^{i}_{\rm c}$ and so on for the two 
remaining species, i.e. $\theta_{\rm b}$ and $\theta_{\nu}$.}:
\begin{equation}
k^2 ( {\cal H} \phi + \psi') = \frac{3}{2} {\cal H}^2 \biggl[ \frac{4}{3} ( R_{\nu} \theta_{\nu} + ( 1 - R_{\nu} ) \theta_{\gamma}) + 
\frac{{\cal F}_{\rm B}(k)}{4 \pi \sigma a^4 \rho} + \theta_{\rm b} \Omega_{\rm b} + \theta_{\rm c} \Omega_{\rm c} \biggr],
\label{p0ilex}
\end{equation}
where ${\cal F}_{\rm B}(k,\eta)$ is the Fourier transform of the generalized Lorentz force, i.e. 
\begin{equation}
 \vec{\nabla} \cdot [ (\vec{\nabla} \times \vec{B}) \times \vec{B}] = \int d^{3} k  
 {\cal F}_{\rm B}(k) e^{ i \vec{k} \cdot \vec{x}},
\end{equation}
and $ {\cal F}_{\rm B}(k)$ is given by the following convolution  
\begin{equation}
 \int d^{3} p [ ( \vec{k}\cdot \vec{p}) B_{j}(p)B^{j}(|\vec{k} - \vec{p}|) - (k_{i} - p_{i})B_{j}(|\vec{k} - \vec{p}|) B_{j}(|\vec{k} - \vec{p}|)
B^{j}(p)]. 
\end{equation}
Notice that a consistent treatment of fully inhomogeneous magnetic fields implies that 
\begin{equation}
\Omega_{\rm B} \ll 1, ~~~~~~~~~\frac{{\cal F}_{\rm B}}{4\pi k^2  \rho a^4} \ll 1,
\label{closure}
\end{equation}
in order not to over-close the Universe. 
In the ideal MHD limit, 
i.e. $\sigma\to \infty$, the contribution of ${\cal F}_{\rm B}(k)$ disappears from Eq. (\ref{p0ilex}).  

Another important condition on the {\em difference} of the conformally 
Newtonian potentials comes from the $(i\neq j)$ components 
of the perturbed Einstein equations:
\begin{equation}
k^2 (\phi - \psi) = \frac{9}{2}{\cal H}^2  \biggl[- \frac{4}{3} \sigma_{\nu} R_{\nu} - \sigma_{\rm B}(k)\biggr],
\label{pineqjex}
\end{equation}
having defined
\begin{equation}
\sigma_{\rm B}= \frac{\Omega_{\rm B}(k)}{3} - \frac{{\cal F}_{\rm B}(k)}{4 \pi a^4 \rho k^2}.
\label{sigmaB}
\end{equation}
The quantity $\sigma_{\rm B}$ can be interpreted as the anisotropic stress arising thanks to the 
inhomogeneous magnetic field and,
in the force-free case, $\sigma_{\rm B} = \Omega_{\rm B}/3$.
Equation (\ref{pineqjex}) relates the difference of the two 
metric fluctuations to a combination of the neutrino and magnetic 
anisotropic stresses.

Let us finally come to the equation stemming from the 
perturbed $(i=j)$ components of the Einstein equations:
\begin{equation}
\psi'' + ( 2 \psi' + \phi') {\cal H} + (2 {\cal H}' + {\cal H}^2) \phi - \frac{k^2}{3} ( \phi - \psi) = \frac{1}{2} {\cal H}^2 [ ( R_{\nu} \delta_{\nu} + (1 - R_{\nu})) 
\delta_{\gamma} + \Omega_{\rm B}(k)],
\label{pijlex}
\end{equation}
where the contribution of the other species vanishes since both baryons and CDM particles have a vanishing barotropic index, i.e. $w_{\rm b} = w_{\rm c} =0$.

Equations (\ref{p00lex}), (\ref{p0ilex}), (\ref{pineqjex}) and (\ref{pijlex}) should
be supplemented by the evolution equations of the various plasma 
quantities. For doing so, there are two complementary approaches.
In the first approach, one may notice that the perturbed Einstein equations 
only involve the density contrasts and the peculiar velocities 
of the various species, i.e. the monopoles and the dipoles of the 
(perturbed) phase spece distributions. Therefore, the simplest 
approach is to describe the evolution of the different species of the 
plasma through an appropriate set of fluid equations (following 
from the covariant conservation equations for each species).
The exception 
in this statement is represented by the neutrinos whose evolution 
includes also the quadrupole of the neutrino phase space 
distribution, i.e. $\sigma_{\nu}$.  To treat neutrinos one should include, therefore, also 
higher multipoles in the Boltzmann hierarchy.
Alternatively, one may treat all the species through a Boltzmann hierarchy. This
is the second, complementary, approach to the problem of initial conditions.

Proceeding along the improved fluid approach the covariant conservation 
equations for photons and baryons imply, respectively, 
\begin{eqnarray}
&& \theta_{\rm b}' + {\cal H} \theta_{\rm b} = k^2 \phi + \frac{{\cal F}_{\rm B}(k)}{ 4\pi a^4 \rho_{\rm b}}
+ \frac{4}{3} \frac{\Omega_{\gamma}}{\Omega_{\rm b}} a n_{\rm e} 
x_{\rm e} \sigma_{\rm T} ( \theta_{\gamma} - \theta_{\rm b}) =0
\label{thb}\\
&& \delta_{\rm b}' = 3 \psi' - \theta_{\rm b}.
\label{db}
\end{eqnarray}
and 
\begin{eqnarray}
&& \delta_{\gamma}' = - \frac{4}{3} \theta_{\gamma} + 4 \psi',
\label{phot1}\\
&& \theta_{\gamma}' = \frac{k^2}{4} \delta_{\gamma} + k^2 \phi  + a n_{\rm e} \sigma_{\rm T} ( \theta_{\rm b} - \theta_{\gamma} ).
\label{phot2}
\end{eqnarray}
In Eqs. (\ref{thb}) and (\ref{phot2}) Thompson scattering term has been added. 
At early times, baryons and photons are synchronized so well that, to lowest 
order in the tight coupling approximation, $ \sigma_{\rm T} \to \infty$ and 
$\theta_{\gamma} \simeq \theta_{\rm b}$.

The evolution of the CDM particles follows from the fluctuations of the covariant conservation equation whose first-order fluctuation 
leads, in Fourier space, to
\begin{eqnarray}
&&\theta_{\rm c}' + {\cal H} \theta_{c} = k^2 \phi,
\label{CDM1}\\
&& \delta_{\rm c}' = 3 \psi' - \theta_{\rm c}.
\label{CDM2} 
\end{eqnarray}

To describe neutrinos during and after horizon crossing requires a Boltzmann hierarchy for $\delta_{\nu}$, $\theta_{\nu}$ and for 
the higher multipole moments, i.e. $\ell \geq 2$,  of the neutrino phase-space density ${\cal F}_{\nu\ell}$.
With this caveat in mind, after neutrino decoupling, at temperatures of about $1$ MeV, massless neutrinos obey, in Fourier space, 
the following set of equations 
\begin{eqnarray}
&& \delta_{\nu}' = - \frac{4}{3} \theta_{\nu} + 4\psi',
\label{nu1}\\
&& \theta_{\nu}' = \frac{k^2}{4} \delta_{\nu} - k^2 \sigma_{\nu} + k^2 \phi,
\label{nu2}\\
&& \sigma_{\nu}' = \frac{4}{15} \theta_{\nu} - \frac{3}{10} k {\cal F}_{\nu 3},
\label{nu3}
\end{eqnarray}
where $\sigma_{\nu} = {\cal F}_{\nu 2}/2$ is the quadrupole moment of the (perturbed) neutrino phase-space distribution and, as 
introduced above, ${\cal F}_{\nu\ell}$ is the $\ell$-th multipole.

The closed system formed by the perturbed Einstein equations 
and by the fluid equations may be solved. For illustrative purposes, consider 
the magnetized adiabatic mode in the force-free case.
The solution can be obtained as an expansion in $k\tau$ which is small, i.e. 
$k\tau < 1$ when the relevant modes are outside the horizon. In the presence 
of a magnetic field the density contrasts turn out to be 
\begin{eqnarray}
&&\delta_{\rm b, c} = \overline{\delta}_{\rm b, c} - \frac{3}{4} \Omega_{\rm B} 
- \biggl[\frac{69 - 61\,R - 8\,R_{\nu}^2}{60\,\left( 25 + 2\,R_{\nu} \right) }\biggr]\Omega_{\rm B}k^2 \tau^2,
\label{dbOM}\\
&&\delta_{\gamma} = \overline{\delta}_{\gamma} - \Omega_{\rm B} 
+ \biggl[\frac{237 + 152\,R_{\nu} + 16\,R_{\nu}^2}{90\,\left( 25 + 2\,R_{\nu} \right) } \biggr] 
\Omega_{\rm B} k^2 \tau^2,
\label{dgOM}\\
&& \delta_{\nu} = \overline{\delta}_{\nu} -\Omega_{\rm B} 
- \biggl[\frac{375 - 207\,R_{\nu} - 152\,R_{\nu}^2 - 16\,R_{\nu}^3}{90\,R_{\nu}\,\left( 25 + 2\,R_{\nu} \right) } \biggr]
\Omega_{\rm B} k^2 \tau^2,
\label{dnOM}
\end{eqnarray}
where the barred quantities denote the adiabatic mode in the absence of magnetic field.
In particular, for the density contrasts  in the absence of inhomogeneous 
magnetic field we have 
\begin{eqnarray}
&& \overline{\delta}_{\rm b} = 
\overline{\delta}_{\rm c} = - \frac{3}{2} \phi_{0} - \frac{\left( 525 + 188\,R_{\nu} + 16\,R_{\nu}^2 \right) }{60\,\left( 25 + 2\,R_{\nu} \right) }\phi_{0} k^2 \tau^2,
\label{ad3}\\
&& \overline{\delta}_{\gamma} = \overline{\delta}_{\nu} 
= - 2 \phi_{0} - \frac{\left( 525 + 188\,R_{\nu} + 16\,R_{\nu}^2 \right) }{45\,\left( 25 + 2\,R_{\nu} \right) }\phi_{0} k^2 \tau^2,
\label{solde1}
\end{eqnarray}
where $\phi_{0}$ is the constant value of $\phi$ to lowest order in $k\tau$.
While in the absence of magnetic field the adiabaticity condition of Eq. (\ref{ad2}) 
is verified both to lowest and to higher order in $k\tau$, in the presence of 
magnetic field the adiabaticity condition is violated to next-to-leading order.

With similar notations we can also report the evolution for the 
metric fluctuations 
\begin{eqnarray}
&& \phi = \overline{\phi} 
-\biggl[\frac{6 - 8\,R_{\nu} + 2\,R_{\nu}^2}{45\,\left( 25 + 2\,R_{\nu} \right) } \Omega_{\rm B} \biggr] k^2 \tau^2,
\label{phOM}\\
&& \psi = \overline{\psi} 
-\biggl[ \frac{69 - 61\,R_{\nu} - 8\,R_{\nu}^2}{180\,\left( 25 + 2\,R_{\nu} \right) }\Omega_{\rm B}\biggr] k^2 \tau^2,
\label{psOM}
\end{eqnarray}
where, as before, $\overline{\phi}$ and $\overline{\psi}$ denote 
the standard adiabatic solution, i.e. 
\begin{eqnarray}
&& \overline{\phi} = \phi_0  -\frac{\left( 75\, + 14\,\,R_{\nu} - 8\,\,R_{\nu}^2 \right) }{90\,\left( 25 + 2\,R_{\nu} \right) }\phi_0 k^2 \tau^2,
\label{solphi1}\\
&& \overline{\psi} 
= \psi_{0} -\frac{\left( 75 + 79\,R_{\nu} + 8\,R_{\nu}^2 \right) }{90\,\left( 25 + 2\,R_{\nu} \right) }\phi_{0} k^2 \tau^2.
\label{solpsi1}
\end{eqnarray}
As in the standard adiabatic case, $\psi_{0}$ is determined in terms 
of $\phi_0$ by the neutrino anisotropic stress. In the magnetized case the 
relation is modified as 
\begin{equation}
\psi_{0} = \phi_{0} \biggl( 1 + \frac{2}{5} R_{\nu}\biggr) - \frac{\Omega_{\rm B}}{5} ( R_{\nu} -1 ).
\end{equation}
The solution for the velocity fields is finally 
\begin{eqnarray}
&&\theta_{\gamma} = \overline{\theta}_{\gamma} - \biggl[\frac{\Omega_{\rm B}}{4} k^2 \tau
+  \frac{\left(  7 + 8\,R_{\nu} \right)   }{40\,\left( 25 + 2\,R \right)}\Omega_{\rm B}  k^4\tau^3\biggr],
\label{thgOM}\\
&&\theta_{\nu} =     \overline{\theta}_{\nu} 
- \frac{\Omega_{\rm B}}{4} \frac{R_{\gamma}}{R_{\nu}}  k^2 \tau     
- \biggl[\frac{45 - 29\,R_{\nu} - 16\,R_{\nu}^2}{72\,R_{\nu}\,\left( 25 + 2\,R_{\nu} \right) }\Omega_{\rm B}\biggr] k^4 \tau^3,
\label{thnOM}\\
&&\theta_{\rm b} =\theta_{\rm c} = \overline{\theta}_{\rm b}  -
 \biggl[\frac{6 - 8\,R_{\nu} + 2\,R_{\nu}^2}{180\,\left( 25 + 2\,R_{\nu} \right) } \Omega_{\rm B}\biggr]  k^4 \tau^3,
\label{thbOM}
\end{eqnarray}
where, following the same notations used before, the standard adiabatic solution is 
\begin{eqnarray}
&& \overline{\theta}_{\nu} = \frac{\phi_0}{2}k^2 \tau - \frac{\left( 65 + 16\,R_{\nu} \right) }{36\,\left( 25 + 2\,R_{\nu} \right) }\phi_0 k^4 \tau^3,
\label{solthnu1}\\
&& \overline{\theta}_{\rm b} = \frac{\phi_{0}}{2} k^2 \tau -\frac{\left( 75\, + 14\,\,R_{\nu} - 8\,\,R_{\nu}^2 \right) }{360\,\left( 25 + 2\,R_{\nu} \right) }\phi_0 k^4 \tau^3,
\label{solthb1}\\
&&\overline{\theta}_{\rm c} = \frac{\phi_{0}}{2} k^2 \tau -\frac{\left( 75\, + 14\,\,R_{\nu} - 8\,\,R_{\nu}^2 \right) }{360\,\left( 25 + 2\,R_{\nu} \right) }\phi_0 k^4 \tau^3,
\label{solthc1}\\
&& \overline{\theta}_{\gamma} =  \frac{\phi_0}{2}k^2 \tau -\frac{\left( 25 + 8\,R_{\nu} \right) }{20\,\left( 25 + 2\,R_{\nu} \right) }\phi_{0} k^2 \tau^2,
\label{solthgam1}
\end{eqnarray}
Notice that while to leading order all the peculiar velocities of the plasma coincide, to 
next-to-leading order they are different: this occurs already in the standard adiabatic case. 

Since the adiabaticity condition is only realized to lowest order in $k\tau$ it is legitimate to name the solution 
presented in Eqs. (\ref{dbOM})--(\ref{thbOM}) {\em quasi-adiabatic}. The corrections to adiabaticity are of order 
$\Omega_{\rm B} k^2 \tau^2$, so they are suppressed outside the horizon and also, by virtue of Eq. (\ref{closure}),  by $\Omega_{\rm B}$. 
Notice that three regimes emerge naturally. The quasi- adiabatic regime where 
$\Omega_{\rm B}(k) \leq \phi_{0}(k)$, the isocurvature regime
$\Omega_{\rm B}(k) > \phi_{0}(k)$ and the fully 
adiabatic regime, i.e. $\Omega_{\rm B}(k) \ll \phi_{0}(k)$. 

It is useful to recall that the presence of fully inhomogeneous 
magnetic fields changes the qualitative features of the 
tight coupling expansion \cite{magnetizedincon}. As it is 
well known, the idea of the tight-coupling approximation 
is to tailor a systematic expansion in powers of the 
inverse optical depth, i.e. $1/\epsilon'$ in the notations 
employed Eq. (\ref{opdep}). Instead of being interested 
in the brightness perturbations related to Q and U we shall 
be interested in the brightness perturbation related to I, i.e. 
the intensity of the scattered radiation. To derive the evolution 
equations in the tight-coupling approximation for a magnetoactive 
plasma we recall that the equation for $\Delta_{\rm I}$ 
can be written as 
\begin{equation}
\Delta_{\rm I}' + i k\mu \Delta_{\rm I} = ( \psi' - i k \mu \phi) + \tau'\biggl[ - \Delta_{\rm I} + \Delta_{{\rm I}0} +  \mu v_{\rm b} - 
\frac{1}{2} P_{2}(\mu) S_{\rm Q}\biggr],
\label{DI1}
\end{equation}
where we defined $ v_{\rm b} = \theta_{\rm b}/(i k)$ and where 
$S_{\rm Q}$ is the same quantity already introduced in Eq. (\ref{source}).
The equation for $v_{\rm b}$ can be readily obtained from Eq. (\ref{thb}) 
and it is given by 
\begin{equation}
v_{\rm b}' + {\cal H} v_{\rm b} = - i k\phi - \frac{\tau'}{\alpha} [3i \Delta_{{\rm I}1} + v_{\rm b}] - i {\cal W}_{\rm B}(k),
\label{vb0}
\end{equation}
where 
\begin{eqnarray}
&&{\cal W}_{\rm B}(k) =  \frac{{\cal F}_{\rm B}}{4\pi k \rho_{\rm b} a^4 },
\label{W}\\
&& \alpha = \frac{3}{4} \frac{\rho_{\rm b}}{\rho_{\rm \gamma}}.
\label{alpha}
\end{eqnarray}
It should be noticed that even if the magnetic field does not 
appear directly in Eq. (\ref{DI1}), it does appear in Eq. (\ref{vb0}).
This implies that the equation for $\Delta_{\rm I}$ (already 
to zeroth order in the tight-coupling expansion) inherits an 
extra source term that depend on the divergence of the Lorentz 
force. This statement can be verified by expanding, consistently, 
the evolution equations for the brightness perturbations 
in powers of $1/|\epsilon'|$ \cite{magnetizedincon}. 
For instance, to zeroth order in the tight-coupling expansion
the (decoupled) evolution equation for the monopole 
can be writtten as 
\begin{equation}
\overline{\Delta}_{{\rm I}0}''  + \frac{\alpha'}{\alpha + 1} \overline{\Delta}_{{\rm I}0}' + \frac{k^2}{3 (\alpha +1)} \overline{\Delta}_{{\rm I}0} = 
 \biggl[ \psi'' + \frac{\alpha'}{\alpha +1} \psi' - \frac{k^2}{3} \phi \biggr] - \frac{\alpha}{\alpha + 1} \frac{ {\cal F}_{\rm B}}{12\pi \rho_{\rm b} a^4}.
\label{monopole}
\end{equation}
In the limit ${\cal F}_{\rm B} \to 0$ the usual decoupled equation for the evolution of the monopole is clearly recovered.

\subsection{Constraints on fully inhomogeneous fields}

In the following we will refer exclusively to the fully inhomogeneous 
case since the uniform case has been already discussed in Section 
3. In \cite{maxst1} the effects of a fully inhomogeneous 
magnetic field has been considered semi-analytically. The 
analysis of \cite{maxst1} was motivated by the results of
where, in a specific model, quite sizable large-scale magnetic 
fields could be onbtained. In particular, the energy spectrum could be, in that case, quasi-flat, i.e. in the notations of Section 4, $m \simeq - 2.9$. It was 
then noted that a rough limit on the magnetic field intensity 
could be obtained, i.e. $B_{L} \leq 10^{-9}$ where we follow the notations 
of Section 4 and take as comoving scale of regularization the Mpc.  

A semi-analytical analysis of the tensor and vector modes performed in 
\cite{tk1}. The authors 
discussed correctly two opposite regime. In the case $m < -3/2$ 
 the two-point function of the Lorentz force 
is dominated by the large-scale features of the magnetic field. On 
the contrary, if $m > -3/2$ the two-point function of the Lorentz 
force is dominated by the high $k$-modes and hence by diffusion. 
In fact, the power-spectrum of the Lorentz force term
can be estimated as 
\begin{eqnarray}
P(k) \simeq\frac{1}{\pi(2m+3)}
\left[\frac{(2\pi)^{m+3}B^2_L}
{\Gamma\left(\frac{n+3}{2}\right)\rho_{\gamma}}\right]^2
\biggl[ \biggl(\frac{k_{D}}{k_{L}}\biggr)^{2m+3} \biggl(\frac{k}{k_{L}}\biggr)^3+\frac{m}{m+3}
\biggl(\frac{k}{k_{L}}\biggr)^{2n+3}\biggr],
\label{LFS}
\end{eqnarray}
where $k_{D}$ is the diffusion scale. The diffusion scale can be estimated, for 
instance, from the considerations of Ref. \cite{b1} where it was argued 
that the diffusion length-scale is basically determined by the Alfv\'en velocity 
times the Silk length-scale. From Eq. (\ref{LFS}) we can see that the 
second term dominates for $-3 < m < -3/2$ while the first term dominates 
for $m > -3/2$. For $ m = -3/2$ both terms may be of the same order. Notice 
that, in Eq. (\ref{LFS}), $k_{L} = 2\pi/L$ and $L$ is the regularization 
length-scale introduced at the end of the previous Section (see Eq. (\ref{Q0})).
  
 In \cite{tk1} the TT, EE, TE and BB correlations \footnote{See the discussion 
 in Section 3 on the cross power spectra whose general definition 
 is given  in Eq. (\ref{defcross}).}
 for vector and tensor modes have been estimated semi-analytically.
 Using $10\%$ estimates the authors of Ref. \cite{tk1} obtain 
 for $m \sim -3$ (e.g. $m \sim -2.9$)
 a limit $B_{L} \leq 7\times10^{-9}\,\,{\rm G}$. This type of limit 
 becomes more stringent as $m$ increases: for $m =0$ we have 
$B_{\rm L} \leq 10^{-10}\,\,\,\,{\rm G}$, while for $m \simeq 2$ 
we would have $B_{\rm L} \leq 10^{-13}\,\,\,\,{\rm G}$.

 Recently in  \cite{lewis2} the calculation of the vector and tensor 
modes has been pushed to $\ell \sim 2000$. It is clear from the 
reported results \cite{lewis2} 
that for a magnetic field with $B_{L} \sim 10^{-9}\, {\rm G}$
and $m \sim -2.9$ the TT correlations induced by the vector modes 
become of the same order to the TT power spectrum (in the 
absence of magnetic field) for $\ell  \geq 2000$. Expected numerical 
disagreements with the semi-analytical estimates of \cite{tk1} 
have been also found for $\ell \geq 100$. 
 \begin{figure}
\centerline{\epsfxsize = 11cm  \epsffile{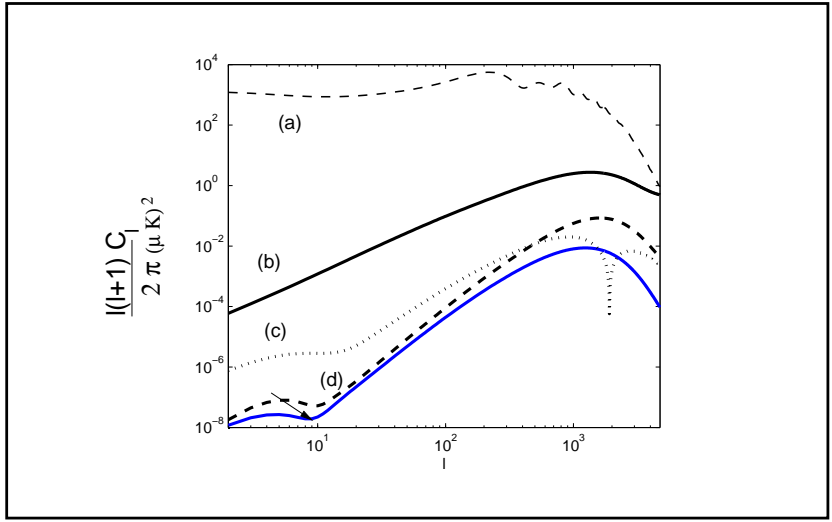}}
\vskip 3mm
\caption[a]{CMB power spectra (vector modes)  for $B_L \sim 3\times 10^{-9}{\rm G}$. 
Adapted from Ref. \cite{lewis1}. TT correlations in the 
absence of magnetic field (curve (a)); TT correlations in the presence of magnetic fields (curve (b)); absolute value of TE correlation 
(curve (c)); BB correlations (curve (d)); EE correlation (indicated by the arrow). These correlations have been defined in Section 3, see, in particular, the 
discussion from Eq. (\ref{defaEB})  to Eq. (\ref{CC1});  the notation EE, BB, TE, and so on, refers to the cross power spectra defined in Eq. (\ref{defcross}) (see also the text).}
\label{FIG5} 
\end{figure}
In Fig. \ref{FIG5} the results obtained in the case of vectors 
for a typical magnetic field $B_L \sim 3\times 10^{-9}{\rm G}$
are reported (see \cite{lewis1}).
 
Scalar modes have been less
studied than the vector and tensor modes. The expectation 
is that scalar modes do not give stronger constraints than vectors or tensors.
It would be, however, desirable to bring the analysis of the scalar problem 
to the same standard of ordinary CMB calculations where the adiabatic mode and the different 
isocurvature modes are constrained to a given confidence level. For instance 
the results reported by \cite{koh} are difficult to intepret since 
bounds on the magnetic field intensity are claimed but the 
initial conditions for the analysis are not clearly specified. 

\subsection{Other effects}

On top on the effects of large-scale magnetic fields on anisotropies 
there are effects directly connected with CMB spectrum like the possible
distorsions.  These have been studied in Ref. \cite{jed} (see also \cite{zizzo}).
The possible distortions of the Planckian spectrum of CMB are 
usually discussed in terms of a chemical potential 
which is bounded, by experimental data, to me $|\mu| < 9\times 10^{-5}$. 
Magnetic field dissipation 
at high red-shift implies the presence of a chemical potential.
Hence bounds on the distorsion of the Planckian spectrum of the CMB can be turned into bounds
on the magnetic field strength at various scales. In particular \cite{jed} the 
obtained bound are such that $B < 3\times 10^{-8}$ G for comoving 
coherence lengths between $0.4$ kpc and $500$ kpc.

As in the case of synchrotron emission also Faraday rotation 
measurements can be used as a diagnostic for foreground 
contamination. The idea would be, in this context to 
look for cross-correlations in the Faraday rotation measure  
of extra-galactic sources and the measured microwave signal 
at the same angular position. 
A recent analysis has been recently reported \cite{coles} 
(see also \cite{bernardi,naselsky1,naselsky2}).

\section{Concluding remarks}
Simple logic dictates that  if the origin of the large-scale magnetic fields is 
 primordial (as opposed to astrophysical) it is plausible to 
  expect the presence of magnetic fields in the primeval plasma also {\em before }
 the decoupling of radiation from matter. As it was argued in the 
 present topical review, CMB anisotropies
 are germane to several aspect of large-scale magnetization. 
 CMB physics may be the tool that will finally enable us 
either to confirm or to rule out the primordial nature  of galactic and clusters 
 magnetic fields seeds. 
 In the next five to ten years the forthcoming  CMB precision polarization experiments 
 will be sensitive in, various frequency channels between above100 GHz and, roughly below 900 GHz. The observations will be conducted both via satellites (like the  Planck satellite) 
 and via ground based detectors (like in the case of the QUIET arrays).
 In a complementary view, the SKA telescope will provide full-sky surveys
 of Faraday rotation that may even get close to 20 GHz.
 
 In an optimistic perspective the forthcoming experimental data together 
 with the steady progress in the understanding of 
 the dynamo theory will  
 hopefully explain the rationale for the ubiquitous nature of large-scale 
 magnetization.  In a pessimistic perspective, the primordial nature 
 of magnetic seeds will neither be confirmed nor ruled out. 
 The plausibility of (highly speculative) theoretical scenarios for the origin of primordial magnetic fields cannot be decided only using the 
 fertile imagination of theoretical physicists. It would 
 be wiser, at the same time, to adopt a model-independent approach by 
 sharpening those theoretical tools that 
 may allow, in the near future, a direct observational test of the effects 
 of large-scale magnetic fields on CMB anisotropies.


\newpage

  
\begin{thebibliography}{99}
\bibitem{zelm1} Ya. Zeldovich,  Sov. Phys. JETP {\bf 21}, 656 (1965).

\bibitem{zelm2} Ya. Zeldovich and I. Novikov, {\it The Structure and Evolution of the Universe}, (Chicago
University Press, Chicago, 1971), Vol.2.

\bibitem{kal1}  I. M. Khalatnikov, Sov. Phys. JETP {\bf 21}, 172 ( 1965).

\bibitem{lif1} E. M. Lifshitz and I. M. Khalatnikov, Sov. Phys. JETP {\bf 12}, 108 (1960).

\bibitem{lif2} E. M. Lifshitz and I. M. Khalatnikov, Sov. Phys. JETP {\bf 12}, 558 (1961).

\bibitem{BK1} V. A. Belinskii and I. M. Khalatnikov, Sov. Phys. JETP {\bf 30}, 1174 (1970).

\bibitem{RS} M.P. Ryan, and L.C. Shepley, {\it Homogeneous Relativistic Cosmologies}, 
( Princeton University Press, Princeton 1975).

\bibitem{unif1} J.~D.~Barrow,  Phys.\ Rev.\ D {\bf 55}, 7451 (1997).

\bibitem{lim1}  J.~D.~Barrow, P.~G.~Ferreira and J.~Silk, Phys.\ Rev.\ Lett.\  {\bf 78}, 3610 (1997).

\bibitem{giofully} M.~Giovannini,  Phys.\ Rev.\ D {\bf 59}, 123518 (1999).

\bibitem{bama} J.~D.~Barrow and R.~Maartens,   Phys.\ Rev.\ D {\bf 59}, 043502 (1999).

\bibitem{giohom1} M.~Giovannini,  Phys.\ Rev.\ D {\bf 62}, 067301 (2000).

\bibitem{bronikov} K.~A.~Bronnikov, E.~N.~Chudaeva and G.~N.~Shikin, Class.\ Quant.\ Grav.\  {\bf 21}, 3389 (2004).

\bibitem{chen} G. Chen {\it et al.},  Astrophys. J. {\bf 611}, 655 (2004).

\bibitem{yates} R. Durrer, T. Kahniashvili and A. Yates,  Phys. Rev. D {\bf 58}, 123004 (1998).

\bibitem{b1}  K.~Subramanian and J.~D.~Barrow,  Phys.\ Rev.\ D {\bf 58}, 083502 (1998).

\bibitem{b2}  K.~Subramanian and J.~D.~Barrow, Phys.\ Rev.\ Lett.\  {\bf 81}, 3575 (1998).

\bibitem{b3} T.~R.~Seshadri and K.~Subramanian,  Phys.\ Rev.\ D {\bf 72}, 023004 (2005).

\bibitem{rg} J.Adams, U.Danielsson, D.Grasso, and H. Rubinstein, {\it Phys. Lett. B} {\bf 388}, 253 (1996). 

\bibitem{rgrep}  D.~Grasso and H.~R.~Rubinstein, Phys.\ Rept.\  {\bf 348}, 163 (2001).

\bibitem{vl1} A. Vlasov, Zh. \'Eksp. Teor. Fiz. {\bf 8}, 291 (1938);
J. Phys. {\bf 9} (Moscow), 25 (1945).

\bibitem{vl2} L. D. Landau, J. Phys. (Moscow)  {\bf 10}, 25 (1945).

\bibitem{ll1} E. M. Lifshitz and L. P. Pitaevskii, {\em Physical Kinetics},
(Pergamon Press, Oxford, England, 1980).

\bibitem{biskamp}  D. Biskamp, {\it Non-linear Magnetohydrodynamics}
(Cambridge University Press, Cambridge, 1994).

\bibitem{boyd}   T. J. M Boyd and J. J. Sanderson, {\it The physics of plasmas}, (Cambridge University 
Press, Cambridge, UK, 2003). 

\bibitem{birefringence} M. Giovannini, Phys. Rev. D {\bf 71}, 021301 (2005).

\bibitem{skilling} J. Skilling, Phys. of Fluids {\bf 14}, 2523 (1971).

\bibitem{q2} S.~M.~Carroll, G.~B.~Field and R.~Jackiw, Phys.\ Rev.\ D {\bf 41}, 1231 (1990).

\bibitem{q3} S.~M.~Carroll, Phys.\ Rev.\ Lett.\  {\bf 81}, 3067 (1998).

\bibitem{q5} M.~Giovannini, Phys.\ Rev.\ D {\bf 61}, 063004 (2000).

\bibitem{q6} M.~Giovannini, Phys.\ Rev.\ D {\bf 61}, 063502 (2000).

\bibitem{q7} B.~Feng, H.~Li, M.~z.~Li and X.~m.~Zhang,  Phys.\ Lett.\ B {\bf 620}, 27 (2005).

\bibitem{ssethi} .~K.~Sethi and K.~Subramanian, Mon.\ Not.\ Roy.\ Astron.\ Soc.\  {\bf 356}, 778 (2005).

\bibitem{breas} K.~R.~S.~Balaji, R.~H.~Brandenberger and D.~A.~Easson,
  JCAP {\bf 0312}, 008 (2003).
  
\bibitem{lue}  A.~Lue, L.~M.~Wang and M.~Kamionkowski,  Phys.\ Rev.\ Lett.\  {\bf 83}, 1506 (1999).

\bibitem{lepora} N.~F.~Lepora, arXiv:gr-qc/9812077.

\bibitem{FRT1} A. Kosowsky and A. Loeb, Astrophys. J {\bf 461}, 1 (1996).

\bibitem{FRT2}  M.~Giovannini,  Phys.\ Rev.\ D {\bf 56}, 3198 (1997).

\bibitem{FRT3} D. Harari, J. Hayward and M. Zaldarriaga, {\it Phys. Rev. D} {\bf 55},  1841 (1997).

\bibitem{cmbrev} M.~Giovannini, Int.\ J.\ Mod.\ Phys.\ D {\bf 14}, 363 (2005).

\bibitem{sakurai} J. J. Sakurai, {\it Modern Quantum Mechanics}, (Addison-Wesley, New York, 1985).

\bibitem{MB} C.-P. Ma and E. Bertschinger, Astrophys. J. {\bf 455}, 7 (1995).

\bibitem{HS1} W.~Hu and N.~Sugiyama, Astrophys.\ J.\  {\bf 444}, 489 (1995).

\bibitem{wmap1} H. V. Peris {\it et al}.,  Astrophys. J. Suppl. {\bf 148}, 161 (2003).

\bibitem{wmap2} A. Kogut {\it et al}., Astrophys. J. Suppl. {\bf 148}, 213 (2003).

\bibitem{scocco} C. Scoccola, D. Harari, and S. Mollerach, Phys. Rev. D {\bf 70}, 063003 (2004).

\bibitem{dolgov1} L.~Campanelli, A.~D.~Dolgov, M.~Giannotti and F.~L.~Villante,  Astrophys.\ J.\  {\bf 616}, 1 (2004).

\bibitem{dolgov2} A.~D.~Dolgov, {\it Lectures given at the International School on Astrophysics 'Daniel Chalonge'} 
( 8th Paris Cosmology Colloquium 2004 and Topical Meeting WMAP and the Early Universe, Paris, France, 9-10 Dec 2004),
  arXiv:astro-ph/0503447.

\bibitem{tk3} A.~Kosowsky, T.~Kahniashvili, G.~Lavrelashvili and B.~Ratra,
  Phys.\ Rev.\ D {\bf 71}, 043006 (2005).

\bibitem{planck} see, for instance, http://www.rssd.esa.int/index.php?project=PLANCK.
  
\bibitem{boomerang} see, for instance, http://oberon.roma1.infn.it/boomerang/b2k/.

\bibitem{masi}  S.~Masi {\it et al.}, arXiv:astro-ph/0507509.

\bibitem{quiet} see, for instance, http://cfcpwork.uchicago.edu/kicp-projects/quiet/.
  
\bibitem{SKA} see, for instance, http://www.skads-eu.org/.
  
 \bibitem{govoni} F.~Govoni and L.~Feretti, Int.\ J.\ Mod.\ Phys.\ D {\bf 13}, 1549 (2004).
 
 \bibitem{feretti} B.~M.~Gaensler, R.~Beck and L.~Feretti, New Astron.\ Rev.\  {\bf 48}, 1003 (2004).
  
\bibitem{krall} N. A. Krall and A. W. Trivelpiece, {\it Principles of
Plasma Physics}, (San Francisco Press, San Francisco 1986).

\bibitem{son}  D.~T.~Son, Phys.\ Rev.\ D {\bf 59}, 063008 (1999).

\bibitem{olesen1} A. Brandenburg, K. Enqvist and P. Olesen   Phys. Rev. D {\bf 54}, 1291  (1996). 

\bibitem{olesen2} A. Brandenburg, K. Enqvist and P. Olesen  Phys. Lett. B {\bf 392} , 395 (1997).

\bibitem{olesen3}  P. Olesen,   Phys. Lett. B {\bf 398}, 321 (1997). 

\bibitem{enqvist1} K. Enqvist, Int. J. Mod. Phys. D {\bf 7}, 331 (1998).

\bibitem{review2} M.~Giovannini, Int.\ J.\ Mod.\ Phys.\ D {\bf 13}, 391 (2004).

\bibitem{widrow} L.~M.~Widrow, Rev.\ Mod.\ Phys.\  {\bf 74}, 775 (2003).

\bibitem{pogosian} L. Pogosian, T. Vachaspati, and S. Winitzki {\it  Phys.Rev.D} {\bf 65}, 083502 (2002).  


\bibitem{q4} M.~Giovannini and M.~E.~Shaposhnikov, Phys.\ Rev.\ Lett.\  {\bf 80}, 22 (1998).

\bibitem{q4a}  M.~Giovannini and M.~E.~Shaposhnikov,  Phys.\ Rev.\ D {\bf 57}, 2186 (1998).

\bibitem{tk2} T.~Kahniashvili and B.~Ratra, Phys.\ Rev.\ D {\bf 71}, 103006 (2005).

\bibitem{abr} M. Abramowitz and I. A. Stegun, {\it Handbook of 
Mathematical Functions} (Dover, New York, 1972).

\bibitem{grad}  I. S. Gradshteyn and I. M. Ryzhik,  
{\it Tables of Integrals, Series and Products (fifth edition)},  (Academic Press, New York, 1994).

\bibitem{maxmis} M.~Giovannini and M.~E.~Shaposhnikov, Phys.\ Rev.\ D {\bf 62}, 103512 (2000).

\bibitem{tk1}  A.~Mack, T.~Kahniashvili and A.~Kosowsky,  Phys.\ Rev.\ D {\bf 65}, 123004 (2002).

\bibitem{PT1}  K. Kajantie, {\it et al}.  Phys. Rev. Lett.  {\bf 77}, 2887 (1996).

\bibitem{kv} T. Kibble and A. Vilenkin, Phys. Rev. D {\bf 52}, 679 (1995).

\bibitem{aho} J. Ahonen and K. Enqvist, Phys. Rev. D {\bf 57}, 664 (1997).

\bibitem{baym} G. Baym, D. Bodeker and L. McLerran,  Phys. Rev. D {\bf 53}, 662 (1996).

\bibitem{spergel} J. Quashnock, A. Loeb and D. Spergel, Astrophys. J. Lett. {\bf 344}, L49 (1989).

\bibitem{cheng1} B. Cheng and A. Olinto,  Phys. Rev. D {\bf 50}, 2421 (1994).

\bibitem{sigl2} G. Sigl, A. Olinto, and K. Jedamzik,  Phys. Rev. D{\bf 55}, 4852 (1997).

\bibitem{dolg2} A. D. Dolgov, hep-ph/0110293.


\bibitem{maxint} M.~Giovannini, Phys.\ Rev.\ D {\bf 62}, 123505 (2000).

\bibitem{mgvariation} M. Giovannini,  Phys.\ Rev.\ D {\bf 64}, 061301 (2001).

\bibitem{bamba} K.~Bamba and J.~Yokoyama, Phys.\ Rev.\ D {\bf 70}, 083508 (2004).

\bibitem{orfeu}  O.~Bertolami and R.~Monteiro, Phys.\ Rev.\ D {\bf 71}, 123525 (2005).

\bibitem{kunze} K.~E.~Kunze,   Phys.\ Lett.\ B {\bf 623}, 1 (2005).

\bibitem{maxst}  M.~Gasperini, M.~Giovannini and G.~Veneziano, Phys.\ Rev.\ Lett.\  {\bf 75}, 3796 (1995).

\bibitem{hec1} D.~Boyanovsky, M.~Simionato and H.~J.~de Vega, Phys.\ Rev.\ D {\bf 67}, 023502 (2003).

\bibitem{hec2} D.~Boyanovsky, H.~J.~de Vega and M.~Simionato,  Phys.\ Rev.\ D {\bf 67}, 123505 (2003).

\bibitem{hec3} D.~Boyanovsky and H.~J.~de Vega,  arXiv:astro-ph/0502212.

\bibitem{kari} K.~Enqvist, A.~Jokinen and A.~Mazumdar, JCAP {\bf 0411}, 001 (2004).

\bibitem{laz1} A. Lazarian, Astron. Astrophys. {\bf 264}, 326 (1992).

\bibitem{rees1} M. J. Rees,  Lect. Notes Phys. {\bf 664}, 1 (2005).

\bibitem{lazarian} A. Lazarian, E. Vishniac and J. Cho, Astrophys. J. {\bf 603}, 180 
(2004); Lect. Notes Phys.  {\bf 614}, 376 (2003).

\bibitem{brandenburg} A.~Brandenburg and K.~Subramanian,
  Phys.\ Rept.\  {\bf 417}, 1 (2005).

 \bibitem{kulsrud1} R. Kulsrud, Annu. Rev. Astron. Astrophys. {\bf 37}, 37 (1999).

\bibitem{kulsrud2} R. Kulsrud and S. Anderson,  Astrophys. J. {\bf 396}, 606
(1992).

\bibitem{robert} I.~Brown and R.~Crittenden, Phys.\ Rev.\ D {\bf 72}, 063002 (2005).

\bibitem{bardeen} J. M. Bardeen, Phys. Rev. D {\bf 22}, 1882 (1980).

\bibitem{tsagas1} C.~G.~Tsagas and J.~D.~Barrow, Class.\ Quant.\ Grav.\  {\bf 14}, 2539 (1997).

\bibitem{tsagas2} C.~Tsagas and R.~Maartens,  Phys.\ Rev.\ D {\bf 61}, 083519 (2000).

\bibitem{tsagas3}  C.~G.~Tsagas, Class.\ Quant.\ Grav.\  {\bf 22}, 393 (2005)

\bibitem{mgknot} M. Giovannini,  Phys. Rev. D {\bf 58}, 124027 (1998). 

\bibitem{tk4} T.~Kahniashvili, G.~Gogoberidze and B.~Ratra, arXiv:astro-ph/0505628.

\bibitem{tsagas4} C.~G.~Tsagas, Class.\ Quant.\ Grav.\  {\bf 19}, 3709 (2002).
  
\bibitem{bat} T.~J.~Battefeld and R.~Brandenberger,  Phys.\ Rev.\ D {\bf 70}, 121302 (2004).

\bibitem{mgvect1}  M.~Giovannini,  Phys.\ Rev.\ D {\bf 70}, 103509 (2004).

\bibitem{mgvect2} M.~Giovannini, Class.\ Quant.\ Grav.\  {\bf 22}, 363 (2005).

\bibitem{jkan1} K .~Subramanian and J.~D.~Barrow,  Mon.\ Not.\ Roy.\ Astron.\ Soc.\  {\bf 335}, L57 (2002).

\bibitem{jkan2}  K.~Subramanian, T.~R.~Seshadri and J.~D.~Barrow,  Mon.\ Not.\ Roy.\ Astron.\ Soc.\  {\bf 344}, L31 (2003).

\bibitem{lewis1} A.~Lewis, Phys.\ Rev.\ D {\bf 70}, 043011 (2004).

\bibitem{lewis2} A.~Lewis,  Phys.\ Rev.\ D {\bf 70}, 043518 (2004).

\bibitem{mgentr}  M.~Giovannini,  Phys.\ Lett.\ B {\bf 622}, 349 (2005).

\bibitem{magnetizedincon} M. Giovannini, Phys. Rev. D {\bf 70}, 123507 (2004).
 
\bibitem{koh} S.~Koh and C.~H.~Lee,  Phys.\ Rev.\ D {\bf 62},083509 (2000). 
 
\bibitem{maxst1} M.~Gasperini, M.~Giovannini and G.~Veneziano,
  Phys.\ Rev.\ D {\bf 52}, 6651 (1995).
  
\bibitem{tsagas5} C.~A.~Clarkson, A.~A.~Coley, R.~Maartens and C.~G.~Tsagas,
  Class.\ Quant.\ Grav.\  {\bf 20}, 1519 (2003).
 
\bibitem{jed}  K.~Jedamzik, V.~Katalinic and A.~V.~Olinto,  Phys.\ Rev.\ Lett.\  {\bf 85}, 700 (2000).
 
\bibitem{zizzo}  A.~Zizzo and C.~Burigana, arXiv:astro-ph/0505259.
 
\bibitem{coles}  P.~Dineen and P.~Coles,  Mon.\ Not.\ Roy.\ Astron.\ Soc.\  {\bf 347}, 52 (2004).
 
\bibitem{bernardi}  G.~Bernardi, E.~Carretti, R.~Fabbri, C.~Sbarra and S.~Cortiglioni,
  arXiv:astro-ph/0508278.
 
\bibitem{naselsky1} P.~D.~Naselsky, A.~G.~Doroshkevich and O.~V.~Verkhodanov,
  Astrophys.\ J.\  {\bf 599}, L53 (2003).
  
\bibitem{naselsky2}   P.~D.~Naselsky, L-Y. Chiang, P. Olesen, O. Verkhodanov, 
Astrophys. J. {\bf 615}, 45 (2004).

\end{thebibliography}
\end{document}